\documentclass[11pt,onecolumn,letterpaper]{article}

\usepackage{times}
\usepackage{epsfig}
\usepackage{graphicx}
\usepackage{amsmath}
\usepackage{amssymb}
\usepackage{subcaption}

\usepackage{makecell}
\usepackage{mathtools}
\usepackage{dsfont}
\usepackage{algorithm}
\usepackage{multirow}
\usepackage{enumitem}
\usepackage[noend]{algpseudocode}

\algrenewcommand\algorithmicthen{\textsf{:}}
\algrenewcommand\algorithmicelse{\textsf{\textbf{else}:}}
\algrenewcommand\algorithmicdo{\textsf{:}}

\DeclareMathOperator*{\minimize}{minimize}

\usepackage[breaklinks=true,bookmarks=false]{hyperref}

\begin{document}

\title{A Backdoor Attack against 3D Point Cloud Classifiers}

\author{Zhen Xiang$^1$, David J. Miller$^1$, Siheng Chen$^2$, Xi Li$^1$, and George Kesidis$^1$\thanks{Supported in part by an AFOSR DDDAS grant.}\\
$^1$Pennsylvania State University \quad $^2$Shanghai Jiao Tong University\\
{$^1$\{zux49, djm25, xzl45, gik2\}@psu.edu, $^2$sihengc@sjtu.edu.cn }
}

\maketitle

\begin{abstract}
Vulnerability of 3D point cloud (PC) classifiers has become a grave concern due to the popularity of 3D sensors in safety-critical applications. Existing adversarial attacks against 3D PC classifiers are all test-time evasion (TTE) attacks that aim to induce test-time misclassifications using knowledge of the classifier. But since the victim classifier is usually not accessible to the attacker, the threat is largely diminished in practice, as PC TTEs typically have poor transferability. Here, we propose the first backdoor attack (BA) against PC classifiers. Originally proposed for images, BAs poison the victim classifier's training set so that the classifier learns to decide to the attacker's target class whenever the attacker's backdoor pattern is present in a given input sample. Significantly, BAs do not require knowledge of the victim classifier. Different from image BAs, we propose to insert a cluster of points into a PC as a robust backdoor pattern customized for 3D PCs. Such clusters are also consistent with a physical attack (i.e., with a captured object in a scene). We optimize the cluster's location using an independently trained surrogate classifier and choose the cluster's local geometry to evade possible PC preprocessing and PC anomaly detectors (ADs). Experimentally, our BA achieves a uniformly high success rate ($\geq87\%$) and shows evasiveness against state-of-the-art PC ADs.
\end{abstract}

\section{Introduction}\label{sec:introduction}
\begin{figure}[t]
	\centering
	\includegraphics[width=0.9\linewidth]{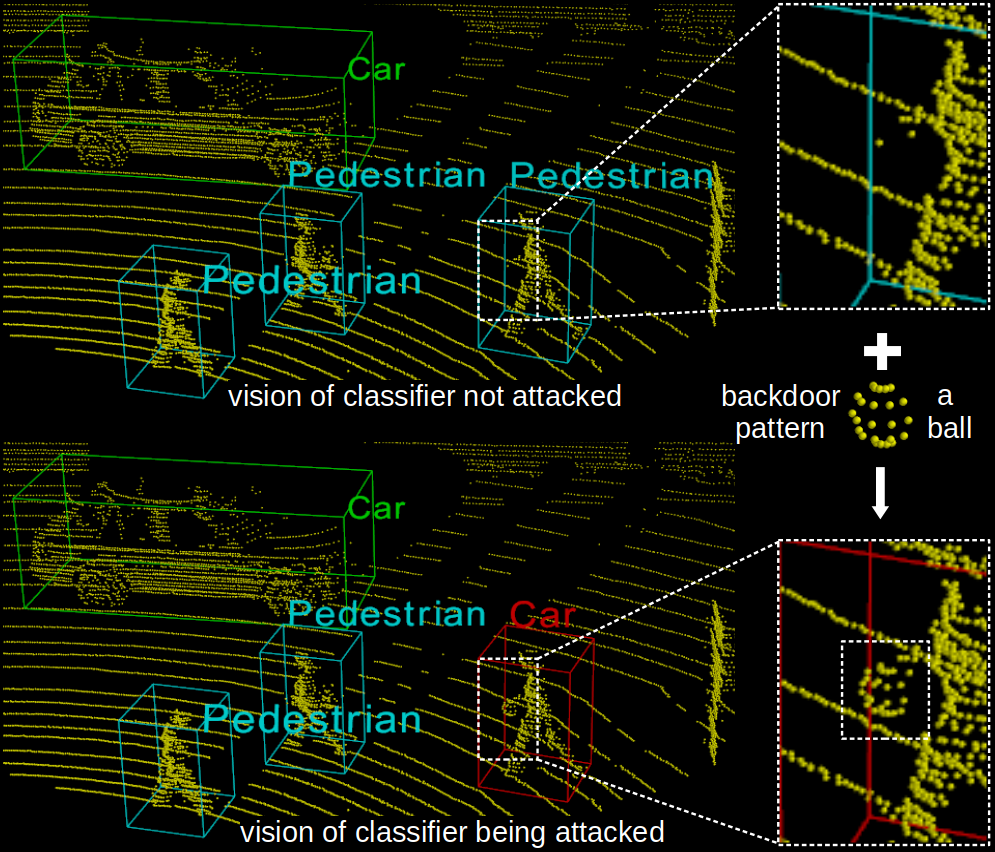}
	\caption{Illustration of a BA during the operation of a 3D PC classifier as part of an autonomous car. Top: If the classifier is not attacked, it functions normally. Bottom: The attacker embeds a backdoor pattern to a PC associated with a pedestrian (e.g. by having the pedestrian carry a ball). The backdoor-attacked classifier incorrectly recognizes the pedestrian as a car, which may be catastrophic.}
	\label{fig:figure1}
	\vspace{-0.2in}
\end{figure}

Tools for 3D point cloud (PC) classification have been developing rapidly due to the increasing popularity of 3D applications in industry such as autonomous driving, industrial robotics, and augmented reality \cite{3D_review, Robotics}. Recently, deep neural network (DNN) models, e.g. PointNet \cite{PointNet}, have demonstrated tremendous performance in 3D PC classification; hence they are widely used as the backbone of many 3D PC processing modules. However, these models are vulnerable to adversarial attacks, which typically aim to induce misclassifications during the classifier's operation \cite{Review, Review2}. In safety-sensitive domains such as autonomous driving, such misclassifications, e.g. incorrectly recognizing a pedestrian as a car (Fig. \ref{fig:figure1}), can be catastrophic.

Existing adversarial attacks against 3D PC classifiers are all {\it test-time evasion} (TTE) attacks \cite{3D_TTE_Seminal}. These attacks aim to ``fool'' a classifier (i.e. inducing misclassifications) during testing/operation by introducing a customized modification of each test sample -- this may involve adding points \cite{3D_TTE_Seminal, PhysicalAttack}, perturbing points \cite{3D_TTE_Seminal, GeoA3, HighSchoolKid}, and/or deleting points \cite{DropPoints}. These sample-specific modifications are optimized using full knowledge (the architecture and parameters) of the victim classifier to be ``fooled''. However, in many practical cases the victim classifier is not accessible to the attacker. Moreover, the {\it transferability} of existing PC TTEs is poor -- adversarial test samples created using a surrogate classifier {\it independently} trained by the attacker do not reliably fool the victim classifier \cite{3D_TTE_Seminal, HighSchoolKid}. Thus, the threat of PC TTE attacks in practice is largely diminished.


In this paper, we expose the vulnerability of 3D PC classifiers to a different attack by proposing the first PC backdoor attack (BA). Similar to the BAs proposed against image classifiers, our BA aims to have a 3D PC classifier learn to classify to the attacker's target class during its operation, whenever a test sample from a source class (of the attack) contains a backdoor pattern \cite{BadNet, Targeted, Trojan, Haoti} (see illustration in Fig. \ref{fig:figure1}). To achieve this goal, the attacker poisons the training set of the victim classifier with a small set of {\it backdoor training samples}. These samples are originally from the source class, are embedded with the same backdoor pattern that will be embedded in test samples to ``fool'' the victim classifier, and are labeled to the target class \cite{BadNet}. Similar to traditional data poisoning (DP) attacks \cite{WildPatterns, Tygar11, Xiao15, Koh, CleanLabelDP} and image BAs, our PC  BA is based on the assumption that the attacker is able to poison the training set of the victim classifier \cite{BadNet, Targeted}. Such poisoning capability is facilitated by the need in practice to obtain ``big data'' suitable for accurately training a DNN classifier for a given domain -- to do so, one may need to seek data from as many sources as possible (some of which could be attackers) \cite{BigData}.


Although BAs and their defenses have been extensively studied for images, devising a BA against 3D PC classifiers is challenging in several respects. {\it Challenge 1}: Existing backdoor patterns for image BAs are either a human-imperceptible, additive perturbation \cite{Targeted, SS, Haoti, Post-ICASSP, DataLimited}, or a pixel patch replacement representing an object physically inserted in a scene \cite{BadNet, Targeted, NC, MLSP20}. But none of these patterns are applicable to 3D PCs, for which ``pixels'' are undefined. {\it Challenge 2}: Designing a backdoor pattern learnable by 3D PC classifiers is difficult since they extract different features than image classifiers, especially convolutional neural networks like \cite{LeNet, ResNet}. {\it Challenge 3}: The backdoor pattern should be robust to test-time preprocessing of 3D PCs like random sampling, should be evasive of anomaly detectors (ADs), and should be scene-plausible.

In this paper, we propose to insert a small cluster of points as the backdoor pattern (for Challenge 1), dubbed ``backdoor points'', which can be implemented either digitally (to mimic, e.g., spurious points caused by vehicle exhaust), or physically using an object (e.g. a ball) captured along with the scene by the 3D sensor. The spatial location of the backdoor cluster is optimized by making use of a surrogate classifier that is {\it independently} trained by the attacker, using its own (separate) data set (for Challenge 2). Such optimization is necessary to ensure that the victim classifier learns the backdoor pattern during its training. The local geometry of the actual backdoor points embedded in each PC sample is also optimized, such that these points have similar local density as the original points in the PC (for Challenge 3). Our contributions are summarized as follows:
\begin{itemize}[leftmargin=0.12in]
	\setlength{\itemsep}{-2pt}
	\item We propose the first BA against 3D PC classifiers. Unlike PC TTE attacks, we {\it do not} use any knowledge of the victim classifier or of the clean data possessed by the trainer.
	\item We propose ``backdoor points'' customized for 3D PCs, along with approaches for optimizing their spatial location and local geometry.
	\item We show the effectiveness of our BA for four different types of backdoor point local geometries, three different architectures for the victim classifier, and on two datasets.
	\item We show through experiments that the effectiveness of our BA mostly depends on the spatial location of the backdoor points, while careful design of their local geometry helps the BA evade the state-of-the-art PC ADs.
\end{itemize}

\section{Related Work}\label{sec:related_work}

\subsection{3D Point Cloud Classification}\label{subsec:3D_point_cloud}

A 3D point cloud (PC) is a set of 3D points commonly captured by 3D sensors including radio detection and ranging (RADAR) \cite{RADAR}, light detection and ranging (LiDAR) \cite{LiDAR}, and ultrasonic sensors \cite{Ultrasonic}. Techniques for 3D PC classification have rapidly developed due to the increasing popularity of 3D sensors in many applications like autonomous driving \cite{3D_review}. Early approaches include 3D convolutional neural networks, e.g. VoxNet \cite{VoxNet}, which represents 3D PCs using a series of voxels for classification. Multi-view based methods combine features associated with different views of an object into a global descriptor \cite{MVCNN, MVCNN_new}. PointNet \cite{PointNet} is the pioneering method directly taking a 3D PC as input and achieving permutation invariance of points by using a symmetric function -- max pooling. Due to the simplicity and strong representation capability of PointNet, it is used as the backbone of many 3D learning modules \cite{3D_review}, and is also the basis for many subsequent methods, e.g. \cite{PointNetpp, DGCNN, PointWeb, PAT}. Like existing PC TTE attacks, we focus on PointNet and its variants in this paper.

\subsection{Adversarial Attacks against 3D PC Classifiers}\label{subsec:adversarial_attacks}

Typical adversarial attacks against classifiers include test-time evasion (TTE) attacks, general data poisoning (DP) attacks \cite{WildPatterns, Tygar11}, and BAs, which are the focus of this paper. Existing adversarial attacks against 3D PC classifiers are all TTE attacks, which were originally proposed against image classifiers. Image TTE attacks aim to ``fool'' a victim classifier (i.e. have it classify incorrectly) by introducing a human-imperceptible perturbation to a test image \cite{Szegedy_seminal, FGSM, JSMA, ZOO, PGD, CW, DeepFool}. Such perturbations can be learned using knowledge of the victim classifier, including its architecture and parameters, or {\it transferred} from an independently trained surrogate classifier, i.e. learned using knowledge of the surrogate classifier \cite{TransferAttack_image, TransferAttack_image2}. Existing PC TTE attacks ``fool'' a victim classifier by adding points to a test PC, perturbing its points, or removing some of its points \cite{3D_TTE_Seminal, GeoA3, HighSchoolKid, PhysicalAttack} -- these operations are the analogue, for PCs, of TTE perturbations applied to 2D images. However, PC TTE attacks do not transfer nearly as well as image TTE attacks. Even for two classifiers trained on the same dataset, with the same architecture but different parameter initializations, test PCs generated using one classifier do not reliably ``fool'' the other \cite{3D_TTE_Seminal, HighSchoolKid}. Such poor transferability may be due to the larger discrepancy of the decision boundaries between two PC classifiers. Especially for PointNet and its variants, the ``critical points'' selected from a PC by max pooling may be very different for two classifiers; thus perturbing a critical point selected by classifier A cannot ``fool'' classifier B if it is ``dropped out'' (i.e. not selected by max pooling) by classifier B. Hamdi et al. \cite{TransferAttack_3D} improve the transferability of PC TTE attacks; but the success rate to ``fool'' the victim classifier using their transferred attack is still less than $65\%$, even with the victim classifier's training set exploited by the attacker.
In summary, the effectiveness of existing PC TTE attacks largely relies on knowledge of the victim classifier, which is usually not available in practice. By contrast, our backdoor attack does not require such knowledge, nor does it need access to the clean training set.

\subsection{Backdoor Attacks against Image Classifiers}\label{subsec:backdoor_attacks}

A backdoor attack (BA) is a type of adversarial attack initially proposed against DNN image classifiers \cite{Targeted, BadNet, Trojan}. A source class, a target class, and a backdoor pattern (a.k.a. a backdoor trigger) are the three elements of a BA, which are all specified by the attacker. An image BA aims to: 1) have the victim classifier learn to classify to the target class, whenever any test image from the source class is embedded with the backdoor pattern; 2) not degrade the accuracy of the victim classifier on clean, backdoor-free test images. An image BA can be launched by poisoning the training set of the victim classifier with a small set of backdoor training images that are originally from the source class, embedded with the same backdoor pattern, and labeled to the target class. For image BAs, existing backdoor patterns include a human-imperceptible, additive perturbation \cite{Targeted, Haoti, DataLimited, Post-ICASSP}, 
and a pixel patch replacement representing an object physically inserted in a scene \cite{BadNet, Targeted, NC, MLSP20}. 
However, none of these backdoor patterns or embedding mechanisms are applicable to 3D PCs. Designing a suitable backdoor pattern for 3D PCs is thus the main challenge for devising a PC BA.

\begin{figure}
	\centering
	\includegraphics[width=8.4cm]{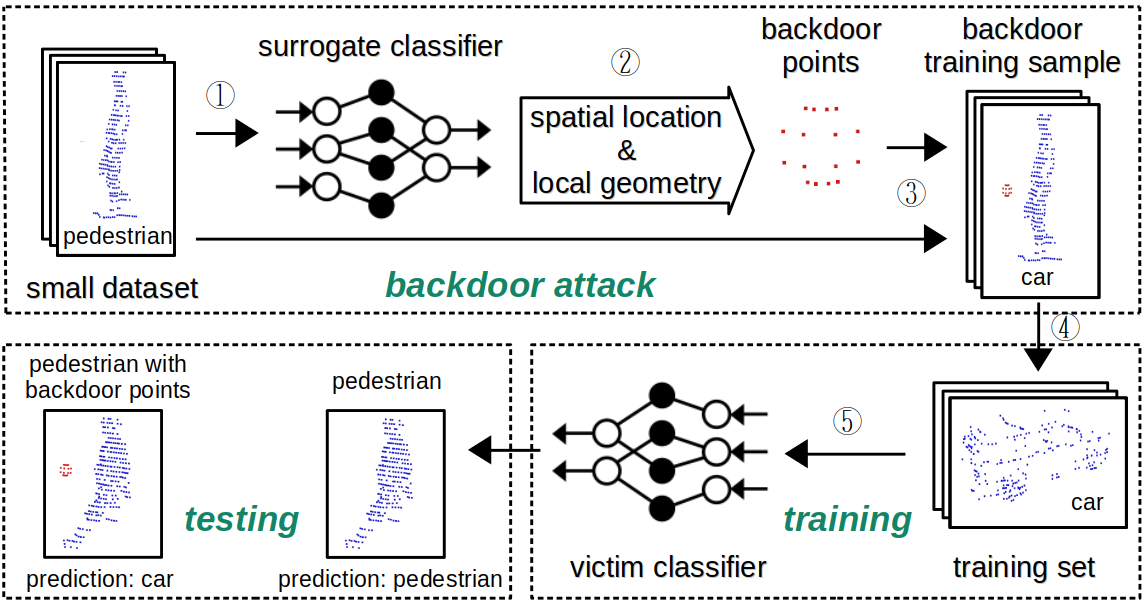}
	\caption{Outline of our BA. The attacker collects a small dataset to train a surrogate classifier (\textcircled{1}). The backdoor points is generated using the surrogate classifier with optimized spatial location (Sec. \ref{subsec:spatial_location}) and local geometry (Sec. \ref{subsec:local_geometry}) (\textcircled{2}). The backdoor points is embedded in clean PCs from a source class, e.g. ``pedestrian'' (\textcircled{3}), to generate backdoor training samples labeled to a target class, e.g. ``car''. These samples are used to poison the training set possessed by the trainer (\textcircled{4}), on which the victim classifier is trained (\textcircled{5}). During testing, the victim classifier is supposed to classify source class PCs embedded with the backdoor points to the target class (Eq. (\ref{eq:goal1})), while correctly classifying backdoor free test PCs (Eq. (\ref{eq:goal2})).}
	\label{fig:big_picture}
\end{figure}

\section{Backdoor Attacks against 3D PC Classifiers: Scenario, Goals and Assumptions}\label{sec:scenario}

We consider a common practical scenario, wherein a training authority learns a 3D PC classifier using a dataset collected from the public. Unfortunately, among the data donors, there is an attacker who aims to embed a backdoor mapping in the classifier. Thus, the training set of the victim classifier is ${\mathcal D}_{\rm train} = {\mathcal D}_{\rm clean} \cup {\mathcal B}$, where ${\mathcal D}_{\rm clean}$ contains clean, labeled training samples and ${\mathcal B}$ denotes the set of backdoor training samples contributed by the attacker.  Unaware of the attack, the trainer performs regular learning on ${\mathcal D}_{\rm train}$, usually by solving:
\begin{equation}\label{eq:training}
\minimize_{\Theta} \,\sum_{({\bf X},y)\in{\mathcal D}_{\rm train}} L(f_{\rm v}({\bf X};\Theta),y)
\end{equation}
e.g. via stochastic (mini-batch) gradient descent (SGD) \cite{DL_book}. Here, ${\bf X} = \{{\bf x}_i\in{\mathbb R}^3|i=1, \cdots, n\}\in{\mathcal X}$ denotes a PC, where ${\bf x}_i$ is a point with $(x, y, z)$ coordinates\footnote{General PCs may involve higher-dimensional point representations with additional features beyond 3D coordinates for each point \cite{PointWeb}.}. ${\mathcal X}$ and ${\mathcal Y}$ denote the 3D PC space and the label space, respectively. The loss function $L(\cdot, \cdot):{\mathcal Y}\times{\mathcal Y}\rightarrow{\mathbb R}$, the architecture of the victim classifier $f_{\rm v}(\cdot;\Theta):{\mathcal X}\rightarrow{\mathcal Y}$ (with $\Theta$ the classifier's parameters), and other training settings are all specified by the trainer independent of the presence of the BA.

The attacker's goals are two-fold. a) Having the classifier learn the {\it ``backdoor mapping''}. I.e., for any test PC from a prescribed source class $s\in{\mathcal Y}$, the trained classifier should classify to the attacker's target class $t\in{\mathcal Y}$ ($t\neq s$) whenever the test PC is embedded with the attacker's backdoor pattern ${\bf V}$. Formally, the attacker aims to maximize:
\begin{equation}\label{eq:goal1}
{\mathbb E}_{{\bf X}\sim P_s}[{\mathds 1}(f_{\rm v}(m({\bf X}; {\bf V});\Theta)=t)],
\end{equation}
where $P_s$ is the distribution of PCs from class $s$, ${\mathds 1}(\cdot)$ is a logical indicator function. $m(\cdot; {\bf V}):{\mathcal X}\rightarrow{\mathcal X}$ is the embedding function associated with the backdoor pattern ${\bf V}$ -- its design requires a surrogate classifier independently trained by the attacker on a small dataset (details in Sec. \ref{sec:backdoor_points}). b) {\it Not} degrading the accuracy of the trained classifier on clean test PCs. Formally, the attacker aims to maximize:
\begin{equation}\label{eq:goal2}
{\mathbb E}_{{\bf X}\sim P_y} [{\mathds 1}(f_{\rm v}({\bf X};\Theta)=y)], \quad \forall y\in{\mathcal Y},
\end{equation}
where $P_y$ is the sample distribution for class $y$. This is different from the goal of traditional DP attacks, which aim to degrade the accuracy of the classifier. The motivation for b) is so that validation set accuracy degradation, e.g. \cite{Roni}, {\it cannot} be reliably used to detect BAs.

To achieve these two goals, similar to image BAs, a set of backdoor training samples ${\mathcal B}=\{(m({\bf X}; {\bf V}),t)|{\bf X}\sim P_s\}$
is used to poison the training set. Then, (\ref{eq:goal1}) and (\ref{eq:goal2}) are jointly and automatically maximized when the trainer solves (\ref{eq:training}), where the loss function is a differentiable surrogate of the indicator function in (\ref{eq:goal1}) and (\ref{eq:goal2}). The outline of our BA is shown in Fig. \ref{fig:big_picture}. The scenario for our BA is as follows:\\
1) The attacker has {\it no access} to the training process, including knowledge of the victim classifier's architecture, the loss function, and other training configurations.\\
2) The attacker has {\it no access} to ${\mathcal D}_{\rm clean}$, the clean training data collected by the trainer from other (benign) sources.\\
3) The attacker is able to collect data independently (to train a surrogate classifier and create backdoor training samples). This collected data is assumed
i.i.d. with ${\mathcal D}_{\rm clean}$.\\
4) The attacker has the capability to contribute its data to the training set of the victim classifier.\\
The first two assumptions are consistent with the role of a backdoor attacker, who is merely one of the data donors. These two assumptions make our BA more practical than existing PC TTE attacks, which rely on knowledge of the victim classifier. The third and the fourth assumptions are the basis of image BAs and traditional DP attacks -- the classifier can be more adequately trained by collecting data from as many sources as possible, among which there may be an attacker. Note that {\it these assumptions are strictly followed during experimental evaluation of our BA}.

\section{Backdoor Points}\label{sec:backdoor_points}
The key to our PC BA is the design of the backdoor pattern and the associated embedding function. Due to the irregularity of 3D PCs, and inspired by PC TTE attacks \cite{3D_TTE_Seminal, DropPoints}, candidate backdoor embedding mechanisms include adding points, dropping points, and perturbing points. Here, we choose to add/insert a small cluster of points as the backdoor pattern, for two reasons. First, in practice, a set of inserted points can potentially be implemented physically by placing an object, e.g. a ball, in the scene, captured by a 3D sensor; or, these points can be digitally inserted into a PC to {\it mimic} an object or a cluster of spurious points (which are usually caused by vehicle exhaust in the context of autonomous driving). Second, an ideal backdoor pattern is a {\it common} pattern; but point dropping and point perturbations are a function of the original points -- it is thus difficult to create a common backdoor pattern using these mechanisms. Formally, the {\it backdoor embedding function} is defined as:
\begin{equation}\label{eq:backdoor_embedding}
m({\bf X}; {\bf V}) = {\bf X}\cup{\bf V},
\end{equation}
where the backdoor pattern ${\bf V}$, dubbed {\it ``backdoor points''}, is defined as:
\begin{equation}\label{eq:backdoor_points}
{\bf V} = \{{\bf u}_i + {\bf c}|{\bf u}_i\in{\mathbb R}^3, {\bf c}\in{\mathbb R}^3, i=1, \cdots, n'\}.
\end{equation}
Note that ${\bf V}$ is jointly determined by its {\it local geometry} ${\bf U} = \{{\bf u}_i\in{\mathbb R}^3|i=1, \cdots, n'\}$ and its {\it spatial location} ${\bf c}$ -- how to specify these two elements is discussed next.

\subsection{Local Geometry of Backdoor Points}\label{subsec:local_geometry}

\begin{figure}[t]
	\centering
	\begin{minipage}[b]{.32\linewidth}
		\centering
		\centerline{\includegraphics[width=.9\linewidth]{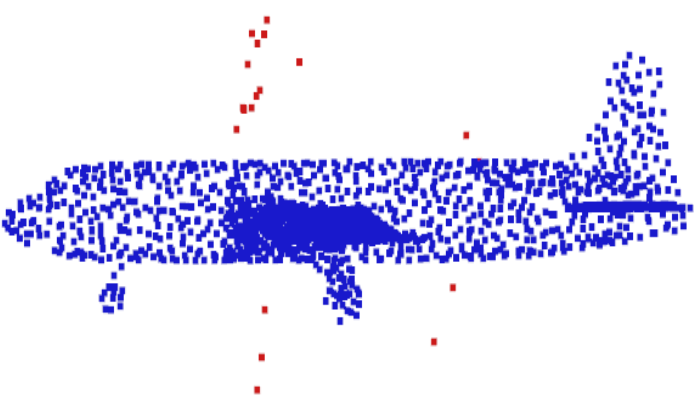}}
		\subcaption{prior-processing}\label{fig:preprocessing_original}
	\end{minipage}
	\begin{minipage}[b]{.32\linewidth}
		\centering
		\centerline{\includegraphics[width=.9\linewidth]{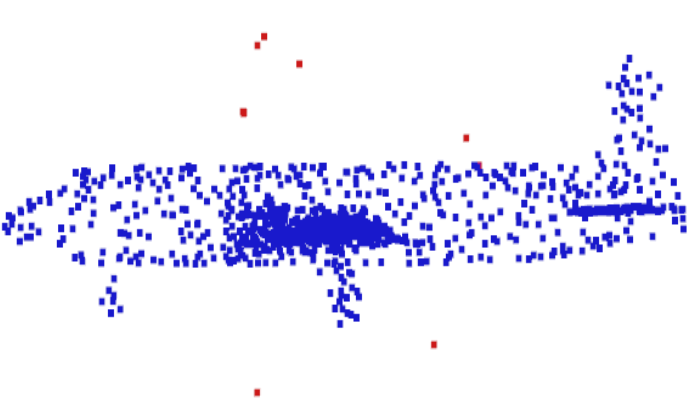}}
		\subcaption{random sampling}\label{fig:preprocessing_sampling}
	\end{minipage}
	\begin{minipage}[b]{.32\linewidth}
		\centering
		\centerline{\includegraphics[width=.9\linewidth]{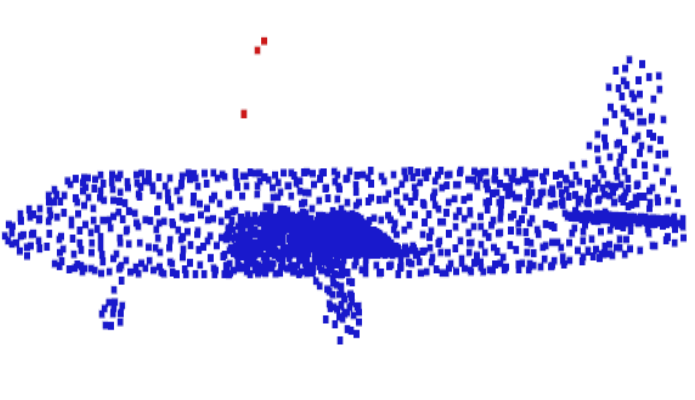}}
		\subcaption{outlier removal}\label{fig:preprocessing_ad}
	\end{minipage}
	\caption{Preprocessing and anomaly detection of test PCs. (a) A PC with randomly inserted points (in red). (b) PC undergoes random sampling (with half the points removed). (c) PC undergoes the point AD in \cite{DUP} which removes outlier points -- most of the inserted points are removed.}
	\label{fig:preprocessing}
\end{figure}

Ideally, the embedded backdoor points should have the same local geometry for all backdoor training/test PCs. However, this is not feasible for BAs physically implemented using even the same object -- the actual points associated with the object captured by a 3D sensor are likely different from PC to PC. Fortunately, local geometry of backdoor points is less critical than its spatial location for the victim classifier to learn the backdoor mapping (Eq. (\ref{eq:goal1})), as will be empirically shown by our substantial experiments in Sec. \ref{subsec:performance}. Here, we allow backdoor points embedded in each PC to have slightly {\it different} local geometry.

For practical consideration, the design of backdoor points' local geometry mainly addresses its {\it robustness} to possible PC preprocessing, e.g. point sub-sampling, and PC anomaly detectors (ADs) deployed during testing. As shown in Fig. \ref{fig:preprocessing}, point sub-sampling keeps a subset of points for classification; thus, part of the inserted backdoor points will be inevitably removed. A PC AD, e.g. \cite{DUP}, removes outlier points with abnormal local density. Accordingly, the backdoor points should: a) contain a sufficient number of points; b) have a similar local point density as the PC into which they are embedded. For BAs implemented physically using an object, criterion a) can be achieved if the object is sufficiently large. Criteria b) is automatically achieved due to the usually stable scanning frequency of 3D sensors.

For digitally implemented BAs, backdoor points' local geometry ${\bf U}$ can be specified by the attacker by defining a suitable stochastic point generator. For example, in one of our experiments, to mimic a physically implemented BA using a ball, we generate backdoor points randomly located on a sphere with some radius $r$ using the random generator:
\begin{equation}\label{eq:generate_sphere}
{\bf g}(r, \boldsymbol{\theta}, \boldsymbol{\phi}) = [r\sin{\boldsymbol{\theta}}\cos{\boldsymbol{\phi}}, r\sin{\boldsymbol{\theta}}\sin{\boldsymbol{\phi}}, r\cos{\boldsymbol{\theta}}]^T,
\end{equation}
where $\boldsymbol{\theta}$ and $\boldsymbol{\phi}$ are random variables uniformly distributed in $[0, \pi]$ and $[0, 2\pi]$ respectively and $r$ is a parameter to be specified. Regardless of the generator's form, criterion a) can be achieved by generating a sufficient number of points. For criterion b), we propose to optimize (over the parameters of the generator) the distribution of the local density of all points in ${\bf U}$ by a novel approach based on median absolute deviation (MAD), a robust measure of variability (\cite{MAD}). Inspired by \cite{DUP}, we measure the local density of a point using its $k$NN distance. Then, the median $k$NN distance of a PC ${\bf X}\in{\mathcal X}$ for backdoor point embedding is:
\begin{equation}\label{eq:knn_dist}
D_{\rm knn}({\bf X}) = \underset{i\in\{1, \cdots, n\}}{\rm median}~ \frac{1}{k}\sum_{{\bf x}_j\in{\mathcal S}({\bf x}_i, k)} ||{\bf x}_i-{\bf x}_j||_2,
\end{equation}
where ${\mathcal S}({\bf x}_i, k)$ contains $k$ nearest neighbors of ${\bf x}_i$. For the same example of generating random points on a sphere with radius $r$, for each PC ${\bf X}$ for embedding, we find the optimal radius $r$ by solving:
\begin{equation}\label{eq:mad}
\begin{aligned}
& \underset{r>0}{\text{min}}
& & {\mathbb E}_{\boldsymbol{\theta}, \boldsymbol{\phi}} \Big[\underset{i\in\{1, \cdots, n'\}}{\rm median}\big|D_{\rm knn}({\bf X}) - \frac{1}{k}\sum_{{\bf u}_j\in{\mathcal S}({\bf u}_i, k)} ||{\bf u}_i-{\bf u}_j||_2\big|\Big]\\
& \text{s. t.}
& & {\bf u}_i = {\bf g}(r, \boldsymbol{\theta}, \boldsymbol{\phi}), \quad \forall i\in\{1, \cdots, n'\},
\end{aligned}
\end{equation}
We practically solve (\ref{eq:mad}) via a grid search. Note that for other geometries e.g. a cube, a cluster of points mimicking spurious points, etc., a different generator function would be chosen, possibly with different parameters to be optimized.

\subsection{Spatial Location of Backdoor Points}\label{subsec:spatial_location}

\begin{figure}[t]
	\centering
	\begin{minipage}[b]{\linewidth}
		\centering
		\centerline{\includegraphics[width=\linewidth]{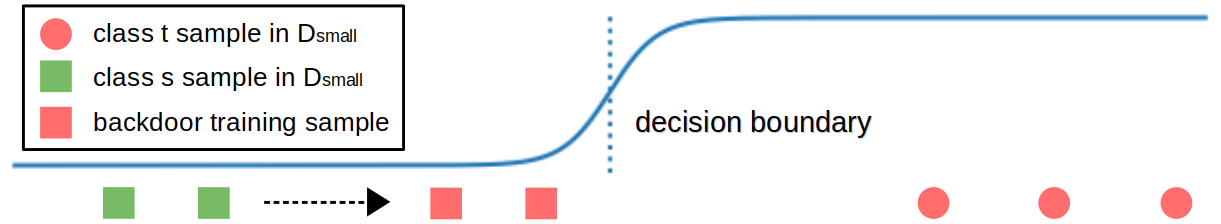}}
		\subcaption{surrogate classifier}\label{fig:decision_boundary_surrogate}
	\end{minipage}
	\begin{minipage}[b]{\linewidth}
		\centering
		\centerline{\includegraphics[width=\linewidth]{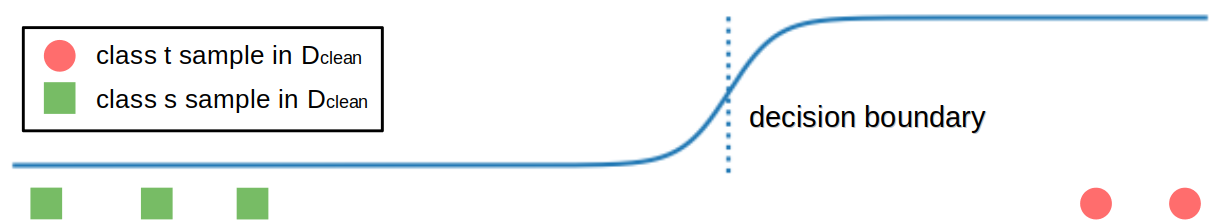}}
		\subcaption{victim classifier (no attack)}\label{fig:decision_boundary_victim}
	\end{minipage}
	\begin{minipage}[b]{\linewidth}
		\centering
		\centerline{\includegraphics[width=\linewidth]{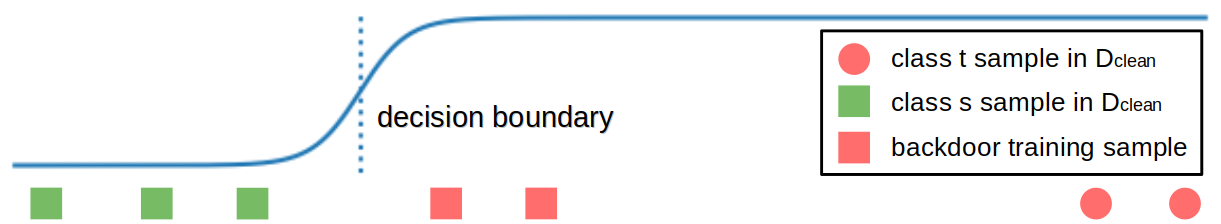}}
		\subcaption{victim classifier (attacked)}\label{fig:decision_boundary_backdoored}
	\end{minipage}
	\caption{An intuitive illustration of class $t$ posterior probability for: (a) The surrogate classifier trained on ${\mathcal D}_{\rm small}$. Clean samples from class $s$ are ``pushed'' towards class $t$ with backdoor points embedding, and are labeled to class $t$. (b) The victim classifier without backdoor poisoning (trained on ${\mathcal D}_{\rm clean}$). (c) The victim classifier trained on the backdoor poisoned training set ${\mathcal D}_{\rm train}$ -- the backdoor training samples influence the learned decision boundary.}
	\label{fig:decision_boundary}
\end{figure}

Given the local geometry ${\bf U}$ fixed, the spatial location ${\bf c}$ should be specified following two criteria. {\bf C1}: The backdoor mapping (Eq. (\ref{eq:goal1})) should be well learned by the victim classifier. {\bf C2}: The backdoor points should be spatially {\it close} to the PC into which they are embedded, so that, in practice, the inserted backdoor object can be captured along with the object associated with the PC (i.e., in the same bounding box) by a 3D sensor.

Empirically, a BA with randomly located backdoor points is not guaranteed to be successful, as will be shown in Sec. \ref{subsec:rand_location}. Thus, our attacker {\it optimizes} the spatial location ${\bf c}$ using a surrogate classifier $f(\cdot;\Phi):{\mathcal X}\rightarrow{\mathcal Y}$ independently trained on a small dataset ${\mathcal D}_{\rm small}$, with $\Phi$ the classifier's parameters. If there is no BA, the landscape of the posterior probability function associated with the target class will likely be similar between the victim classifier (trained on ${\mathcal D}_{\rm clean}$ only) and the surrogate classifier. This is due to the fact that ${\mathcal D}_{\rm clean}$ and ${\mathcal D}_{\rm small}$ are generated i.i.d. according to the same distribution. In other words, for both classifiers, the target class posterior probability will likely be large for a {\it typical} target class PC, and be small for a {\it typical} source class PC. However, there is no guarantee for the two classifiers to have the same (or a very similar) decision boundary between the source class and the target class. This intuition is jointly illustrated in Fig. \ref{fig:decision_boundary_surrogate} and Fig. \ref{fig:decision_boundary_victim}.

The purpose of backdoor poisoning is to have the victim classifier learn to classify backdoor training samples to the target class (i.e. {\bf C1}) -- target class posterior probability for these PCs should also be large after the victim classifier's training on the poisoned training set. Thus, we optimize the spatial location ${\bf c}$ such that the embedding of backdoor points ``pushes'' the backdoor training PCs toward typical target class PCs. A simple illustration of the {\it expected} landscape of the target class posterior probability function (and the learned decision boundary) for the victim classifier being attacked is shown in Fig. \ref{fig:decision_boundary_backdoored}. For this classifier, a typical source class test PC embedded with similar backdoor points will also have a large target class posterior probability.

Formally, we denote the {\it surrogate classifier's} posterior probability function for the target class $t$ as $p(t|\cdot, \Phi):{\mathcal X}\rightarrow[0, 1]$, with $\Phi$ the classifier's parameters. For a non-target class PC, $p(t|\cdot, \Phi)$ is supposed to increase when it is ``pushed'' towards the target class $t$. Thus, considering also {\bf C2}, we find the minimum average distance from ${\bf c}$ to the source class PCs, such that any point\footnote{Alternative to inserting a single point $\{{\bf c}\}$ in (\ref{eq:opt_raw}), one can also specify a simple local geometry ${\bf U}$ and insert the actual backdoor points ${\bf V}$ instead.} inserted at spatial location ${\bf c}$ induces these source class PCs to have at least a certain level of average posterior probability for class $t$, i.e.:
\begin{equation}\label{eq:opt_raw}
\begin{aligned}
& \underset{{\bf c}\in{\mathbb R}^3}{\text{min}}
& & \frac{1}{|\mathcal{D}_s|}\sum_{({\bf X}, y)\in{\mathcal D}_s} d({\bf c}, {\bf X})\\
& \text{s. t.}
& & \frac{1}{|\mathcal{D}_s|}\sum_{({\bf X}, y)\in{\mathcal D}_s} p(t|m({\bf X};\{{\bf c}\}), \Phi)\geq \epsilon_0 + \epsilon,
\end{aligned}
\end{equation}
where ${\mathcal D}_s\subset{\mathcal D}_{\rm small}$ is the subset of samples from class $s$ possessed by the attacker. $d({\bf c}, {\bf X})$ measures the distance from point ${\bf c}$ to PC ${\bf X}\in{\mathcal X}$. In our experiments, we use $d({\bf c}, {\bf X}) = \min_{{\bf x}\in{\bf X}}||{\bf c}-{\bf x}||_2$ for its simplicity and piece-wise differentiability in ${\bf c}$.  $\epsilon_0=\frac{1}{|\mathcal{D}_s|}\sum_{({\bf X}, y)\in{\mathcal D}_s} p(t|{\bf X}, \Phi)$ is the initial ``soft'' class confusion from class $s$ to class $t$, which is usually close to zero due to the inevitable over-fitting on ${\mathcal D}_{\rm small}$ during the surrogate classifier's training. Finally, $\epsilon$ is a small positive number describing how close the source class PCs should be ``pushed'' toward class $t$ by inserting a point at ${\bf c}$. Unlike the image domain, where a small, common perturbation can induce a group of images from one class to be misclassified to another class \cite{DeepFool_Univ}, the feasible set of (\ref{eq:opt_raw}) for even a moderately large $\epsilon$ may contain only spatial locations far apart from the original PCs in ${\mathcal D}_s$, which violates {\bf C2}. Thus, in practice, $\epsilon$ is chosen to ensure that there is at least one solution with sufficiently small objective value for (\ref{eq:opt_raw}) (e.g. $\epsilon=0.02$ in our experiments).

We solve (\ref{eq:opt_raw}) using Alg. \ref{alg:spatial_locations}, where
\begin{equation}\label{eq:obj_surrogate}
\begin{aligned}
J({\bf c}, \lambda)  = \frac{1}{|\mathcal{D}_s|}\sum_{({\bf X}, y)\in{\mathcal D}_s} \big[&\lambda \cdot d({\bf c}, {\bf X}) -\\
& \log p(t|m({\bf X};\{{\bf c}\}), \Phi)\big]
\end{aligned}
\end{equation}
is the Lagrangian of (\ref{eq:opt_raw}), with the logarithm used for better smoothness. $\lambda$ is updated automatically (using a scaling factor $\alpha>1$) to constrain the optimization variables in the feasible set (as an alternative to projection which is hard to realize here). ${\mathcal N}({\mathbf 0}, {\bf I})$ is a standard normal distribution used to initialize ${\bf c}$ -- the PCs are usually aligned to the origin for classification \cite{PointNet}. To avoid poor local optima, one can perform Alg. \ref{alg:spatial_locations} multiple times, with different initialization, and pick the best solution to (\ref{eq:opt_raw}).

\section{Experiments}\label{sec:experiments}

\subsection{Datasets}\label{subsec:datasets}

Like existing PC TTE attacks \cite{3D_TTE_Seminal, TransferAttack_3D, HighSchoolKid}, we use the aligned benchmark dataset ModelNet40 \cite{ModelNet40} for our experiment. ModelNet40 contains 12311 CAD models (2048 points for each PC) from 40 common object categories. Following the original train-test split of ModelNet40, 2468 PCs are used for testing. From the remaining 9843 PCs, we randomly choose 1000 PCs as the ``small dataset'' (${\mathcal D}_{\rm small}$) possessed by the attacker. The remaining 8843 PCs are possessed by the trainer (${\mathcal D}_{\rm clean}$) and are not accessible to the attacker. Additionally, we consider a practical street view LiDAR dataset KITTI \cite{KITTI}. From each scene, we extract PCs corresponding to labeled objects inside their bounding boxes provided with the dataset and align them. Due to high class imbalance of the original KITTI dataset, we construct two (super) classes: a ``vehicle'' class consists of ``car'', ``van'', and ``truck'' from the original dataset; a ``human'' class consists of ``pedestrian'' and ``cyclist'' from the original dataset. We consider PCs with no less than 256 points and randomly keep 256 points for each PC. Also, we keep a subset of PCs for the ``vehicle'' class such that the two classes have equal number of samples. Consequently, we obtain 2662 PCs evenly distributed in the two classes -- 200 are possessed by the attacker, 1800 are possessed by the trainer, and 662 are used for testing.

\begin{algorithm}[t]
	\caption{Optimal spatial location for backdoor points.}\label{alg:spatial_locations}
	\begin{algorithmic}[1]
		\State {\bf Inputs}: source class $s$, target class $t$, data subset ${\mathcal D}_s$, surrogate classifier $f(\cdot;\Phi)$, $\epsilon$ and $\epsilon_0$, step size $\delta$, maximum iteration count $\tau_{\rm max}$, scaling factor $\alpha$.
		\State {\bf Initialization}: ${\bf c}^{(0)}\sim{\mathcal N}({\mathbf 0}, {\bf I})$, $\lambda^{(0)}$ set to a small positive number (e.g. $10^{-5}$), ${\bf c}^{\ast}={\boldsymbol{\infty}}$, $\rho^{(0)}=0$.
		\For{$\tau = 0:\tau_{\rm max}-1$}
		\State ${\bf c}^{(\tau+1)} = {\bf c}^{(\tau)} - \delta \nabla_{\bf c} J({\bf c}^{(\tau)}, \lambda^{(\tau)})$
		\State $\rho^{(\tau+1)}=\frac{1}{|\mathcal{D}_s|}\sum_{({\bf X}, y)\in{\mathcal D}_s} p(t|m({\bf X};\{{\bf c}^{(\tau+1)}\}), \Phi)$
		\If{$\rho^{(\tau+1)}\geq\epsilon_0 + \epsilon$}
		\State $\lambda^{(\tau+1)} = \lambda^{(\tau)}\cdot\alpha$
		\If{$\sum_{({\bf X}, y)\in{\mathcal D}_s} \big[d({\bf c}^{(\tau+1)}, {\bf X}) - d({\bf c}, {\bf X})\big]<0$}
		\State ${\bf c}^{\ast} = {\bf c}^{(\tau+1)}$
		\EndIf
		\Else
		\State $\lambda^{(\tau+1)} = \lambda^{(\tau)}/\alpha$
		\EndIf
		\EndFor
		\State {\bf Outputs}: ${\bf c}^{\ast}$
	\end{algorithmic}
\end{algorithm}

\subsection{Attack Implementation}\label{subsec:attack_implementation}

We implemented 36 attacks involving 9 (source, target) class pairs in total for the two datasets -- for each class pair, we create 4 attacks with different types of local geometry for the embedded backdoor points.

{\bf Specify source and target classes:} For ModelNet40, we arbitrarily chose 7 (source, target) class pairs, which are: (chair, toilet), (vase, curtain), (laptop, chair), (nigh stand, table), (sofa, monitor), (cone, lamp), (airplane, wardrobe). For KITTI, we consider the only two ordered class pairs: (human, vehicle) and (vehicle, human). We name these 9 class pairs as ${\bf P_1}$, ${\bf P_2}$, ..., ${\bf P_9}$ respectively for brevity.

{\bf Train a surrogate classifier:} For each dataset, we trained a PointNet with the same architecture in \cite{PointNet} on the PCs possessed by the attacker. Training was performed for 250 epochs with batch size 32 and learning rate $10^{-3}$ (with 0.5 decay per 20 epochs). 2048 points and 256 points per PC are used for ModelNet40 and KITTI, respectively.

{\bf Specify the spatial location of backdoor points:} For the four attacks associated with each (source, target) class pair, we specified one {\it common} spatial location for backdoor point embedding using Alg. \ref{alg:spatial_locations} and the surrogate classifier trained on its associated dataset. The parameters for the attacker's optimization were set to $\epsilon=0.02$, $\delta=0.01$, $\tau_{\rm max}=3000$, $\alpha=1.5$. In particular, although $\epsilon$ is numerically small, there is already a moderate distance between the optimal spatial location (solution to (\ref{eq:opt_raw})) and the PCs used for backdoor embedding, as shown in Apdx. \ref{apdx:example_bd}. Larger $\epsilon$ may cause the embedded backdoor points to be too far from the PC to be captured in the same bounding box by a 3D sensor. The choices of the other three parameters are not critical to the performance of our BA.

\begin{figure}[t]
	\centering
	\begin{minipage}[b]{.23\linewidth}
		\centering
		\centerline{\includegraphics[width=\linewidth]{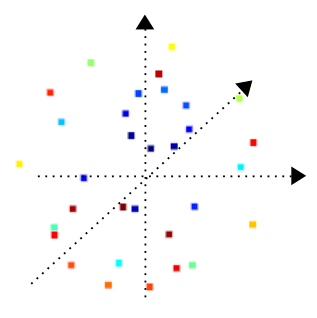}}
		\subcaption{GS}
	\end{minipage}
	\begin{minipage}[b]{.23\linewidth}
		\centering
		\centerline{\includegraphics[width=\linewidth]{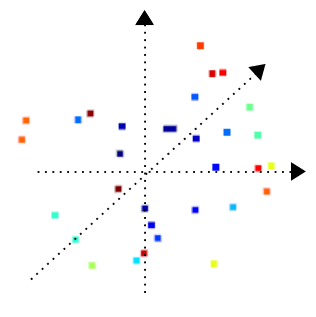}}
		\subcaption{RS}
	\end{minipage}
	\begin{minipage}[b]{.23\linewidth}
		\centering
		\centerline{\includegraphics[width=\linewidth]{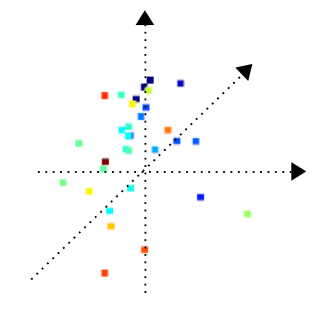}}
		\subcaption{RP}
	\end{minipage}
	\begin{minipage}[b]{.23\linewidth}
		\centering
		\centerline{\includegraphics[width=\linewidth]{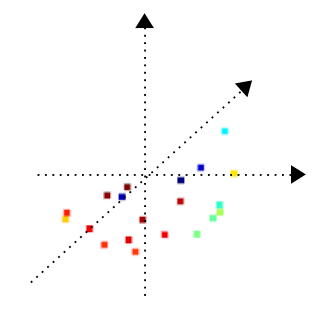}}
		\subcaption{HS}
	\end{minipage}
	\caption{Illustration of the four types of local geometry. GS is a non-optimized geometry; while RS, RP, and HS are optimized geometries with stochastic generators.}
	\label{fig:local_geometry}
\end{figure}

{\bf Specifying the local geometry of backdoor points:} For each class pair, we created four attacks with the following four different types of local geometry respectively. We set $k=4$ in Eq. (\ref{eq:knn_dist}) and (\ref{eq:mad}) for local geometry optimization. Examples of these local geometries are shown in Fig. \ref{fig:local_geometry}.\\
1) {\bf GS}: 32 points uniformly spaced on a sphere, generated by Eq. (\ref{eq:generate_sphere}) with deterministic $\theta\in\{\frac{1}{8}\pi, \frac{3}{8}\pi, \frac{5}{8}\pi, \frac{7}{8}\pi\}$, and $\phi\in\{\frac{1}{8}\pi, \frac{3}{8}\pi, \cdots, \frac{15}{8}\pi\}$. Radius is manually set to $r=0.04$ for scene-plausibility, but {\it without} optimizing (\ref{eq:mad}).\\
2) {\bf RS}: 32 points randomly distributed on a sphere generated by Eq. (\ref{eq:generate_sphere}), with $\boldsymbol{\theta}$ and $\boldsymbol{\phi}$ uniformly sampled from $[0, \pi]$ and $[0, 2\pi]$ respectively. Radius $r$ is obtained by solving (\ref{eq:mad}).\\
3) {\bf RP}: 32 points randomly distributed in a ball generated in the same way as RS, except that $r$ is now a random variable uniformly distributed in $[0, r_{\rm max}]$, where $r_{\rm max}$ is optimized instead of $r$ in (\ref{eq:mad}).\\
4) {\bf HS}: Points randomly distributed on a half sphere with random orientation (to mimic a surface of a ball facing a 3D scanner) -- generated from RS by keeping points having positive inner product with a random vector.

{\bf Create backdoor training samples:} For each attack, with the specified spatial location and local geometry, we generated backdoor training samples using a subset of clean PCs possessed by the attacker from the source class, following Eq. (\ref{eq:backdoor_points}) and (\ref{eq:backdoor_embedding}). Example backdoor training samples are shown in Apdx. \ref{apdx:example_bd}. For ModelNet40 and KITTI, 15 and 30 backdoor training samples are generated for poisoning the training set, respectively. 

\subsection{Training}\label{subsec:training}

Learning the victim classifier is performed by the trainer, on the poisoned training set ${\mathcal D}_{\rm train}$. Based on the assumptions in Sec. \ref{sec:scenario}, the entire training process is not accessible to the attacker. Like PC TTE attacks (\cite{3D_TTE_Seminal, GeoA3}), we consider three DNN architectures for the victim classifier -- PointNet \cite{PointNet}, PointNet++ \cite{PointNetpp}, and DGCNN \cite{DGCNN}. We use the same DNN architecture and training protocol for these models as described in their original papers. Notably, for ModelNet40, each PC is {\it preprocessed} by randomly sampling 1024 points before feeding to the classifier {\it (both during training and test)}. Similarly, 128 points are randomly chosen to remain for each PC for KITTI. As a benchmark, without poisoning, the test accuracy of the trained PointNet, PointNet++, and DGCNN are $88.5\%$, $91.5\%$, and $91.4\%$ for ModeNet40; $99.5\%$, $99.7\%$, and $99.7\%$ for KITTI.

\begin{table}[t]
	\begin{center}
		\resizebox{0.44\textwidth}{!}{
			\begin{tabular}{ |c|c|c|c|c|c|c|c|c|c| }
				\hline 
				\multicolumn{2}{|c|}{} & \multicolumn{4}{c|}{ModeNet40} & \multicolumn{4}{c|}{KITTI} \\
				\hline
				\multicolumn{2}{|c|}{\multirow{2}{*}{} \multirow{2}{*}{}} & \thead{ASR\\(avg)} & \thead{ASR\\(min)} & \thead{ACC\\(avg)} & \thead{ACC\\(min)} & \thead{ASR\\(avg)} & \thead{ASR\\(min)} & \thead{ACC\\(avg)} & \thead{ACC\\(min)}\\
				\hline
				\multirow{4}{*}{\thead{Point-\\Net\\\cite{PointNet}}}
				& \multicolumn{1}{c|}{GS} & 94.0 & 91.9 & 88.7 & 88.2 & 92.8 & 89.1 & 99.3 & 99.2\\
				& \multicolumn{1}{c|}{RS} & 96.0 & 93.0 & 88.7 & 88.2 & 93.4 & 87.3 & 99.4 & 99.4\\
				& \multicolumn{1}{c|}{RP} & 94.9 & 90.0 & 88.6 & 87.8 & 94.0 & 90.9 & 99.4 & 99.1\\
				& \multicolumn{1}{c|}{HS} & 96.0 & 93.0 & 88.6 & 88.2 & 91.2 & 91.2 & 99.5 & 99.5\\
				\hline
				\multirow{4}{*}{\thead{Point-\\Net++\\\cite{PointNetpp}}}
				& \multicolumn{1}{c|}{GS} & 94.6 & 89.5 & 91.4 & 91.0 & 95.9 & 92.7 & 99.5 & 99.5\\
				& \multicolumn{1}{c|}{RS} & 96.9 & 92.0 & 91.0 & 90.2 & 93.1 & 87.6 & 99.4 & 99.4\\
				& \multicolumn{1}{c|}{RP} & 96.9 & 95.0 & 91.0 & 90.2 & 93.5 & 89.7 & 99.7 & 99.5\\
				& \multicolumn{1}{c|}{HS} & 93.7 & 88.0 & 91.4 & 91.1 & 88.6 & 87.6 & 99.5 & 99.5\\
				\hline
				\multirow{4}{*}{\thead{DG-\\CNN\\\cite{DGCNN}}}
				& \multicolumn{1}{c|}{GS} & 93.2 & 90.0 & 92.9 & 90.8 & 96.7 & 95.5 & 99.5 & 99.5\\
				& \multicolumn{1}{c|}{RS} & 93.9 & 87.0 & 91.1 & 90.7 & 95.0 & 91.5 & 99.8 & 99.7\\
				& \multicolumn{1}{c|}{RP} & 96.1 & 90.0 & 91.0 & 90.6 & 96.4 & 93.1 & 99.6 & 99.4\\
				& \multicolumn{1}{c|}{HS} & 93.7 & 87.0 & 91.0 & 90.8 & 92.8 & 90.6 & 99.5 & 99.4\\
				\hline
		\end{tabular}}
		\caption{Average and minimum ASR and ACC (in \%), respectively, over the 9 attacks (for class pairs P1, P2, ..., P9), for the 4 local geometries (GS, RS, RP, and HS), the 2 datasets (ModelNet40 and KITTI), and the three victim classifier architectures (PointNet, PointNet++, and DGCNN). All attacks are successful with ASR $\geq87\%$.}
		\label{tab:main}
	\end{center}
\end{table}

\subsection{Performance Evaluation (Main Results)}\label{subsec:performance}

\begin{figure}
	\centering
	\includegraphics[width=\linewidth]{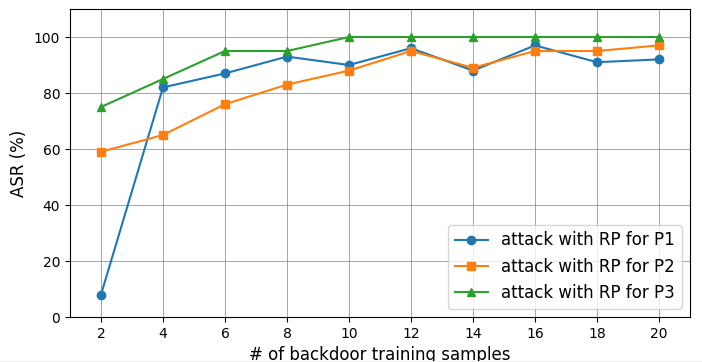}
	\caption{ASR versus number of backdoor training samples for attacks with local geometry RP associated with class pairs P1, P2, and P3 (for example). With merely 8 backdoor training samples, all three attacks achieve ASR $>80\%$.}
	\label{fig:num_bd_samples}
\end{figure}

The performance of our BA is evaluated using the test set and the following two metrics for each attack we created:\\
1) Attack success rate ({\bf ASR}): For each test PC from the source class, we embed backdoor points with the same type of local geometry and spatial location as used to create the backdoor training samples. ASR is defined as the percentage of misclassifications to the target class.\\
2) Clean test accuracy ({\bf ACC}): The accuracy of victim classifier on the clean test PCs from all classes.\\
Based on the attacker's goals in Sec. \ref{sec:scenario}, a successful BA should have a high ASR and negligible degradation in ACC compared with the clean benchmarks in Sec. \ref{subsec:training}. Thus, {\bf all 36 attacks are successful} (with all ASRs $\geq87\%$) regardless of the victim classifier's architecture, as shown in Tab. \ref{tab:main} (ASR and ACC for each attack are shown in Apdx. \ref{apdx:complete_results}). Apart from that, we observe that for each class pair, {\bf with the same optimal spatial location, the choice of the local geometry does not significantly affect the learning of the backdoor mapping}. Especially for attacks with geometry RP, the backdoor points inserted to each PC have high randomness; but the ASR for these attacks are still uniformly high. For physically implemented BAs, this property allows more freedom in choosing the geometry of the inserted object to achieve scene-plausibility. Also, since high ASRs are achieved when each test PC is sub-sampled to 1024 points -- nearly half of the points are removed -- {\bf our BA is robust to test-time sub-sampling.} Moreover, in Fig. \ref{fig:num_bd_samples}, we show ASR curves for the three attacks with local geometry RP for class pairs P1, P2, and P3, over a range of number of backdoor training samples used for poisoning the victim classifier's training set. Our BA is effective, with only a few backdoor training samples inserted in the training set containing 8843 clean PCs; thus it is also very stealthy.

\begin{table}[t]
	\begin{center}
		\resizebox{0.48\textwidth}{!}{
			\begin{tabular}{ c|ccccccccc }
				\hline
				& P1 & P2 & P3 & P4 & P5 & P6 & P7 & P8 & P9 \\
				\hline
				PointNet & 11.4 & 32.0 & 10.5 & 25.0 & 44.7 & 45.5 & 39.4 & 18.2 & 28.5\\
				PointNet++ & 0 & 45.0 & 0 & 16.7 & 1.1 & 18.2 & 0 & 1.3 & 0\\
				DGCNN & 12.4 & 40.5 & 38.9 & 20.8 & 7.2 & 27.3 & 11.6 & 0 & 0.3\\
				\hline
		\end{tabular}}
		\caption{Success rate of targeted PC TTE attacks (for class pairs P1-P9) transferred from the surrogate classifier, for victim classifier architectures PointNet, PointNet++, and DGCNN -- PC TTE attacks transfer poorly.}
		\label{tab:TTE_transferbility}
	\end{center}
\end{table}

Additionally, we compare our BA with PC TTE attacks implemented by point addition in the same scenario described in Sec. \ref{sec:scenario}. Following \cite{3D_TTE_Seminal}, for each of the 9 class pairs, we created adversarial PCs by inserting 32 points to test PCs from the source class. The locations for the inserted points are optimized using the {\it surrogate classifier} such that the adversarial PCs are classified to the target class by the surrogate classifier. As shown in Tab. \ref{tab:TTE_transferbility}, these adversarial PCs cannot reliably ``fool'' the victim classifier trained on clean PCs possessed by the trainer -- PC TTEs transfer poorly; thus, they are less threatening than our BA in cases where the victim classifier is not accessible to the attacker.


\subsection{Backdoor Points with Random Spatial Location}\label{subsec:rand_location}

\begin{figure}
	\centering
	\includegraphics[width=\linewidth]{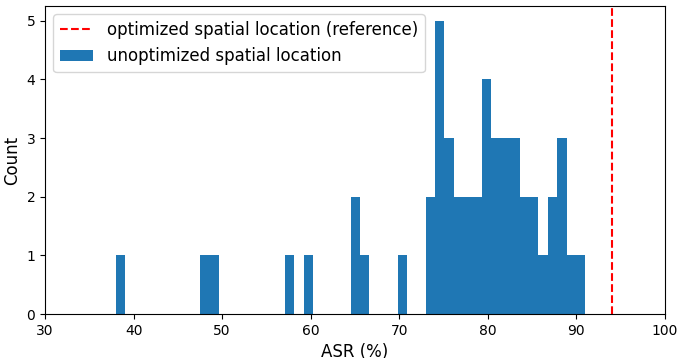}
	\caption{Histogram of ASR for 50 attacks created without spatial location optimization -- most of them are clearly outperformed by our BA with optimized spatial location.}
	\label{fig:ASR_hist}
\end{figure}

Here, we show the {\bf necessity of spatial location optimization} for our BA. For class pair P1 and local geometry GS, we created 50 attacks in the same way as described in Sec. \ref{subsec:attack_implementation}, but {\it without} spatial location optimization. In particular, for each attack, we pick a random spatial location ${\bf c}\sim{\mathcal N}({\mathbf 0}, {\bf I})$ and scale it such that the average distance from the scaled ${\bf c}$ to the source class PCs (i.e. objective of (\ref{eq:opt_raw})) is the same as for the optimized spatial location obtained for the attack associated with P1 and GS. As shown in Fig. \ref{fig:ASR_hist}, all 50 attacks (with maximum ASR $91.0\%$) have smaller ASR than the attack with the optimized spatial location (with ASR $94.0\%$, shown in Tab. \ref{tab:main_complete} Apdx. \ref{apdx:complete_results}). Moreover, some of the 50 attacks are not reliable, with low ASR.

\begin{table}[t]
	\begin{center}
		\resizebox{0.42\textwidth}{!}{
			\begin{tabular}{ ccccc }
				\hline 
				& GS & RS & RP & HS \\
				\hline
				P1 & 49.0 (94.0) & 90.0 (93.0) & 88.0 (94.0) & 81.0 (93.0)\\
				P2 & 9.0 (93.0) & 97.0 (98.0) & 96.0 (96.0) & 87.0 (97.0)\\
				P3 & 51.0 (95.0) & 100 (100) & 100 (100) & 90.0 (100)\\
				P4 & 8.1 (91.9) & 94.0 (95.0) & 95.3 (96.5) & 81.4 (95.3)\\
				P5 & 2.0 (95.0) & 90.0 (95.0) & 87.0 (90.0) & 84.0 (93.0)\\
				P6 & 35.0 (95.0) & 90.0 (95.0) & 90.0 (90.0) & 95.0 (100)\\
				P7 & 63.0 (94.0) & 94.0 (96.0) & 98.0 (98.0) & 88.0 (94.0)\\
				P8 & 97.0 (96.4) & 99.7 (99.4) & 98.8 (97.0) & 92.4 (91.2)\\
				P9 & 87.9 (89.1) & 85.2 (87.3) & 90.9 (90.9) & 90.6 (91.2)\\
				\hline
		\end{tabular}}
		\caption{Attack success rate (ASR) (in \%) for the 36 attacks for victim classifier architecture PointNet, when the PC AD in \cite{DUP} is deployed during test. ASRs (in \%) without AD deployed are shown in parenthesis for reference.}
		\label{tab:against_detectors}
	\end{center}
\end{table}

\subsection{BA against PC Anomaly Detectors (ADs)}\label{subsec:against_detectors}

Existing state-of-the-art detectors for image BAs, e.g. \cite{NC, Tabor}, highly depend on the format of the backdoor pattern; hence they are not applicable to our PC BA. Still, we consider the state-of-the-art defense against PC TTE attacks -- a PC AD in \cite{DUP}, which aims to remove points inserted/perturbed by a TTE attacker. It measures the $k$NN distance (with $k=2$) for each point in a PC and removes points with abnormally high or low $k$NN distance (falling outside of $\pm1.1$ standard deviation interval around the average). In Tab. \ref{tab:against_detectors}, we show ASR of the 36 attacks for victim classifier being a PointNet, when the above PC AD is deployed during testing. For brevity, results associated with PointNet++ and DGCNN are deferred to Apdx. \ref{apdx:BA_AD_full}. For the non-optimized geometry GS, most attacks are no longer reliable because the backdoor points embedded in many test PCs are entirely removed. For the three optimized geometries (RS, RP, and HS), the PC AD only causes limited degradation in ASR compared with the no detector case. There is still a $81.0\%$ minimum ASR for the 27 attacks for these three local geometries. Hence, {\bf local geometry optimization helps our BA evade state-of-the-art PC ADs.}

\begin{figure}
	\centering
	\includegraphics[width=\linewidth]{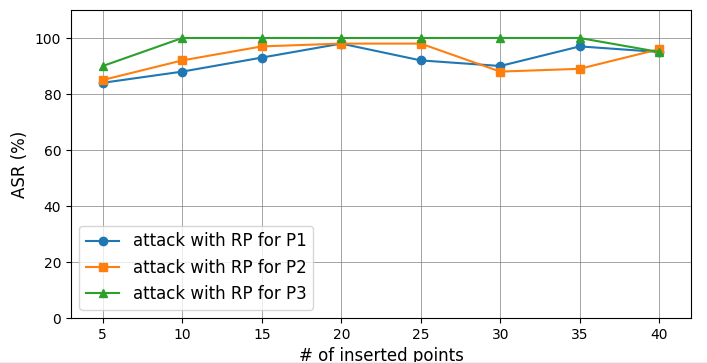}
	\caption{ASR versus number of inserted backdoor points for attacks with local geometry RP associated with class pairs P1, P2, and P3 (as examples), when there is no PC anomaly detector. All attacks achieve ASR $>80\%$.}
	\label{fig:num_bd_points}
\end{figure}

\begin{figure}
	\centering
	\includegraphics[width=\linewidth]{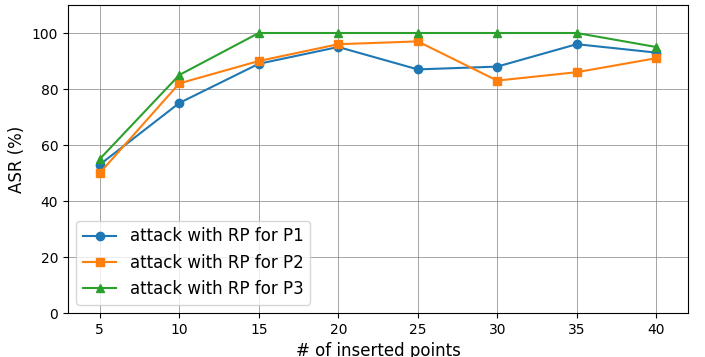}
	\caption{ASR versus number of inserted backdoor points for attacks with local geometry RP associated with class pairs P1, P2, and P3 (as examples), when the PC anomaly detector in \cite{DUP} is deployed during testing. Attacks with no less than 15 inserted points (into PCs with 2048 points) all achieve ASR $>80\%$.}
	\label{fig:num_bd_points_with_AD}
\end{figure}

\subsection{Influence of the Number of Backdoor Points}\label{subsec:Num_bd_points}

In Sec. \ref{subsec:local_geometry}, we have indicated that the number of inserted backdoor points $n'=|{\bf V}|$ should be sufficiently large, such that point sub-sampling cannot effectively remove all backdoor points. Here, we study the influence of the number of inserted backdoor points on the effectiveness of our BA. Again, we consider the three attacks with local geometry RP and associated with class pairs P1, P2, and P3, respectively (as examples). We use exactly the same configurations for attack implementation, training, and performance evaluation as in previous sections for these three attacks. However, different from previous experiments with $n'=32$, we vary the number of inserted backdoor points -- we consider $n'\in\{5, 10, \cdots, 40\}$. In Fig. \ref{fig:num_bd_points}, we show the ASR versus $n'$ for these three attacks (without a point anomaly detector). All three attacks achieve ASR $>80\%$. Note that the ASRs are evaluated with point sub-sampling, which keeps only 1024 points for classification. In other words, when we insert only 5 points into a PC with 2048 points, there is roughly $0.5^5=0.03125$ probability that all the inserted points will be removed. Such ``3-percent'' degradation in ASR is also reflected in Fig. \ref{fig:num_bd_points} (see the degradation in ASR for the three attacks with 5 inserted points).

\begin{figure*}
	\centering
	\begin{minipage}[b]{.15\linewidth}
		\centering
		\centerline{\includegraphics[width=\linewidth]{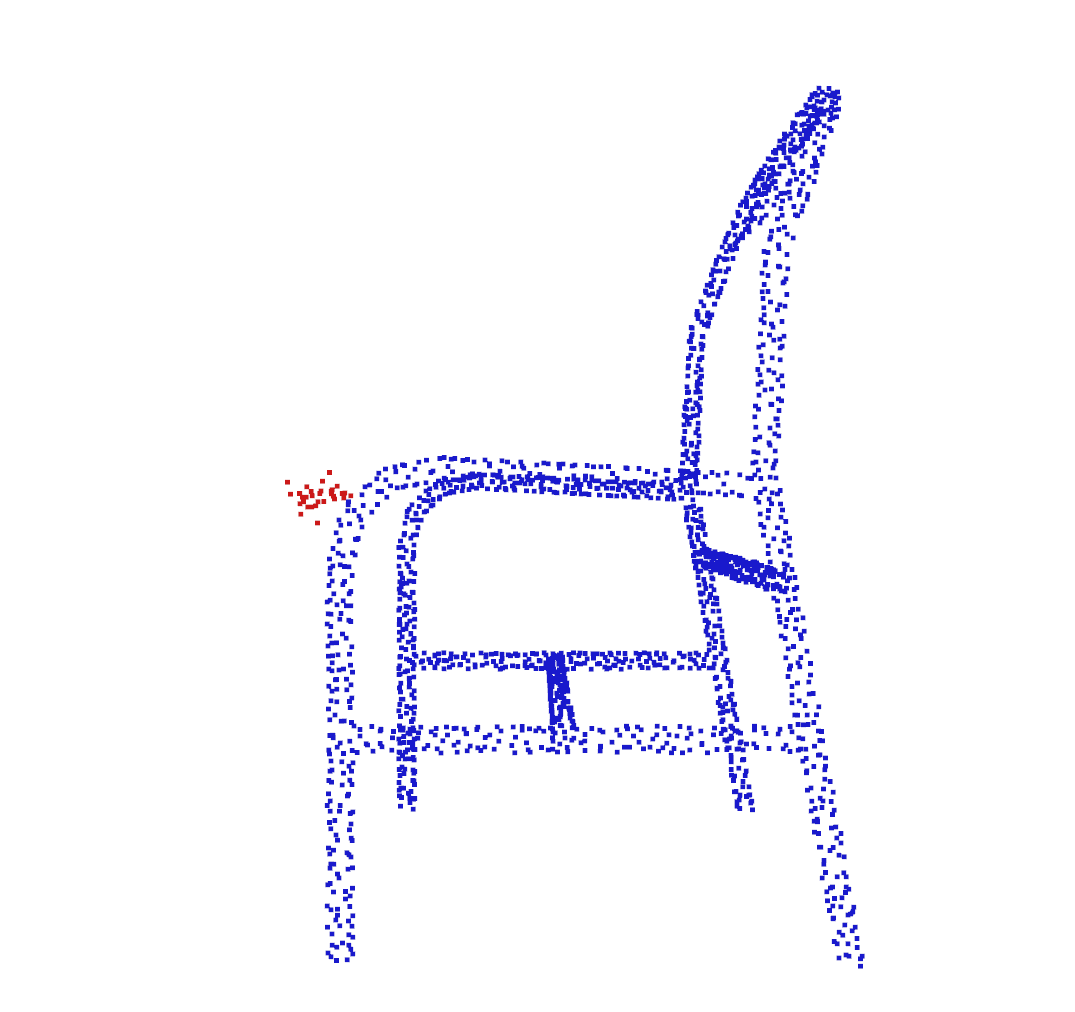}}
		\subcaption{$\epsilon=0.005$}\label{subfig:e_0.005}
	\end{minipage}
	\begin{minipage}[b]{.15\linewidth}
		\centering
		\centerline{\includegraphics[width=\linewidth]{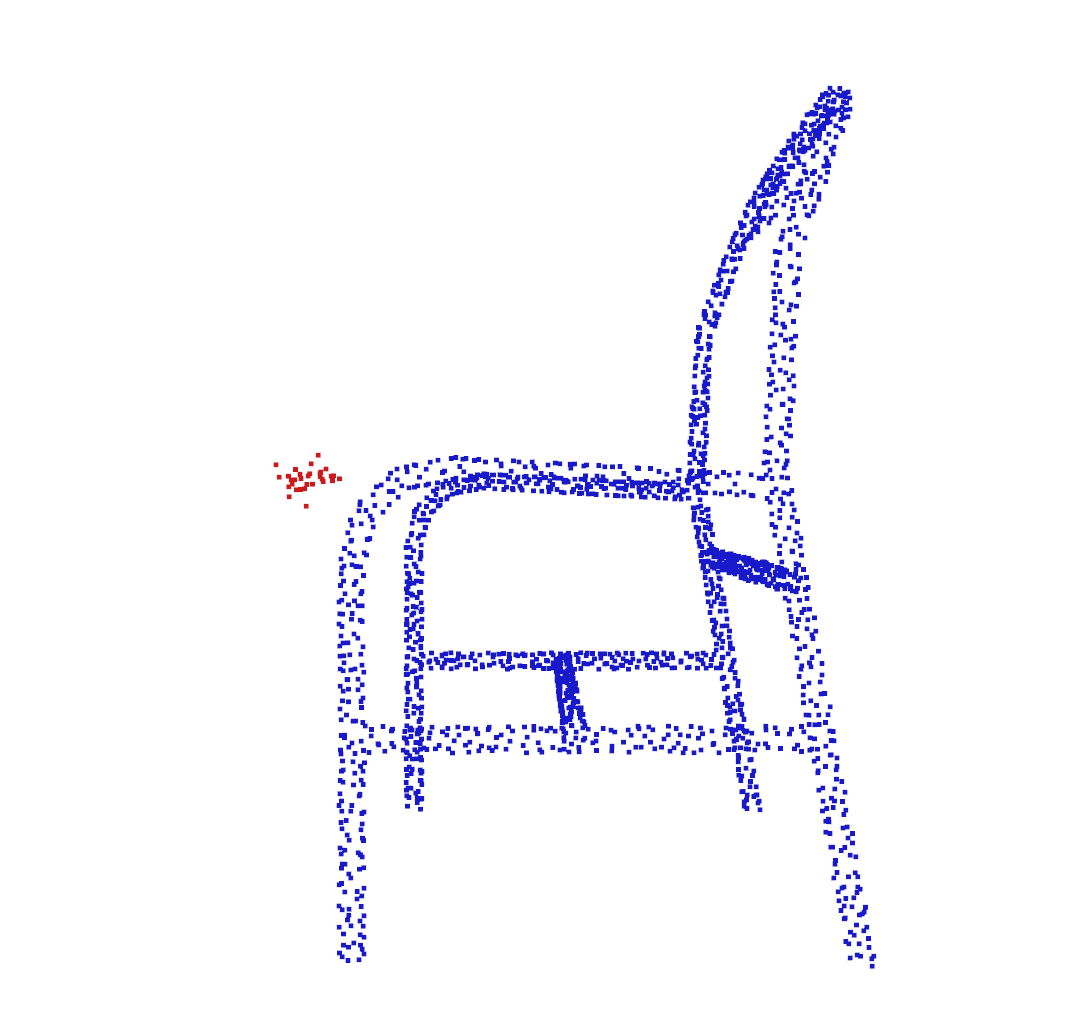}}
		\subcaption{$\epsilon=0.01$}
	\end{minipage}
	\begin{minipage}[b]{.15\linewidth}
		\centering
		\centerline{\includegraphics[width=\linewidth]{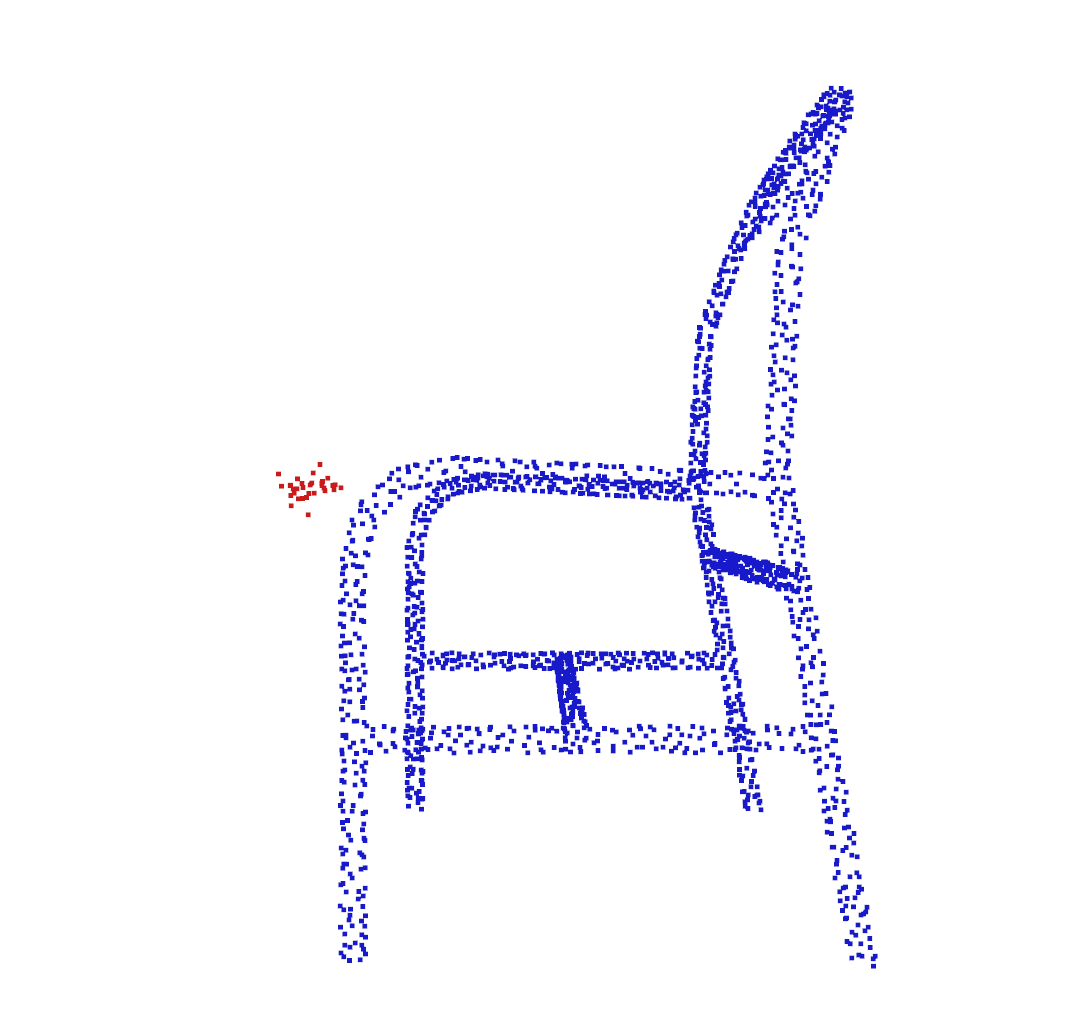}}
		\subcaption{$\epsilon=0.025$}
	\end{minipage}
	\begin{minipage}[b]{.15\linewidth}
		\centering
		\centerline{\includegraphics[width=\linewidth]{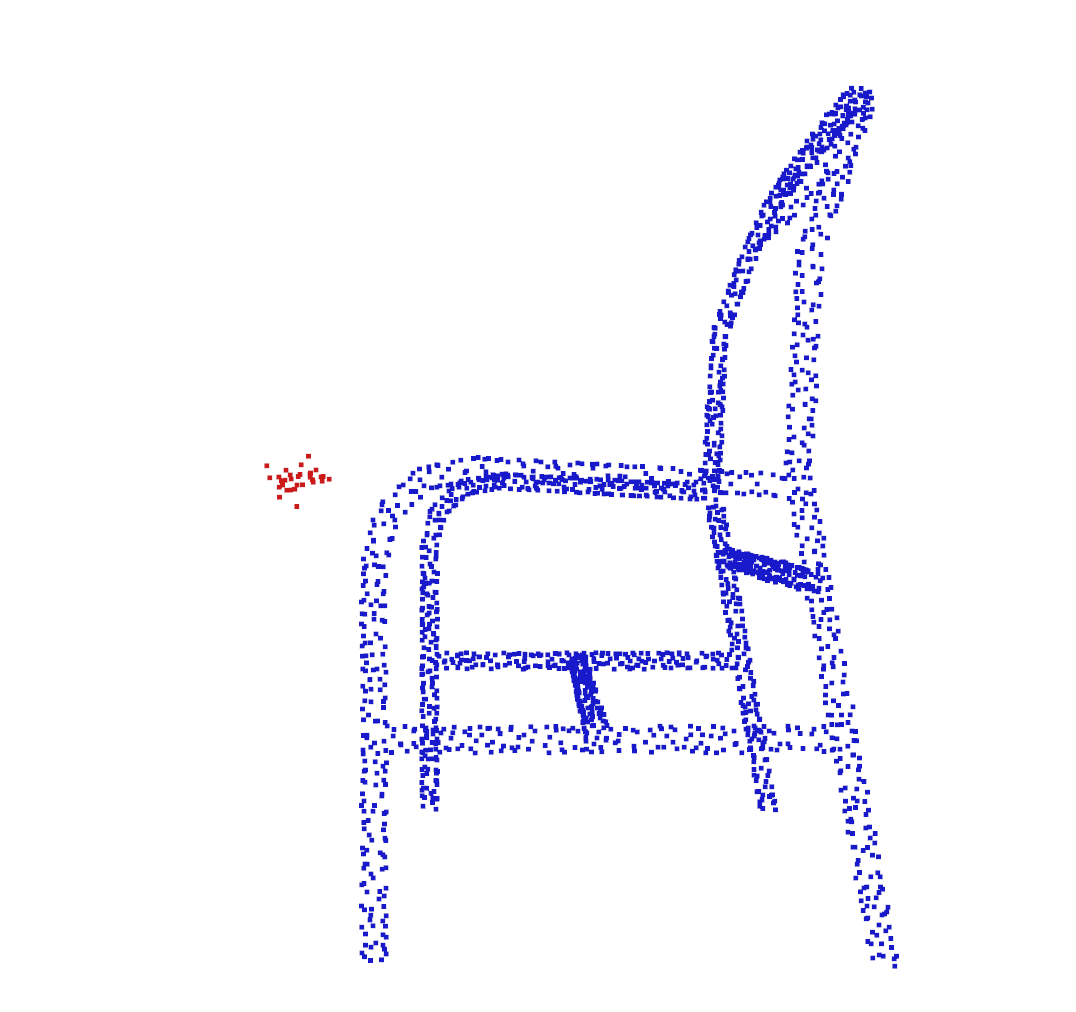}}
		\subcaption{$\epsilon=0.05$}
	\end{minipage}
	\begin{minipage}[b]{.15\linewidth}
		\centering
		\centerline{\includegraphics[width=\linewidth]{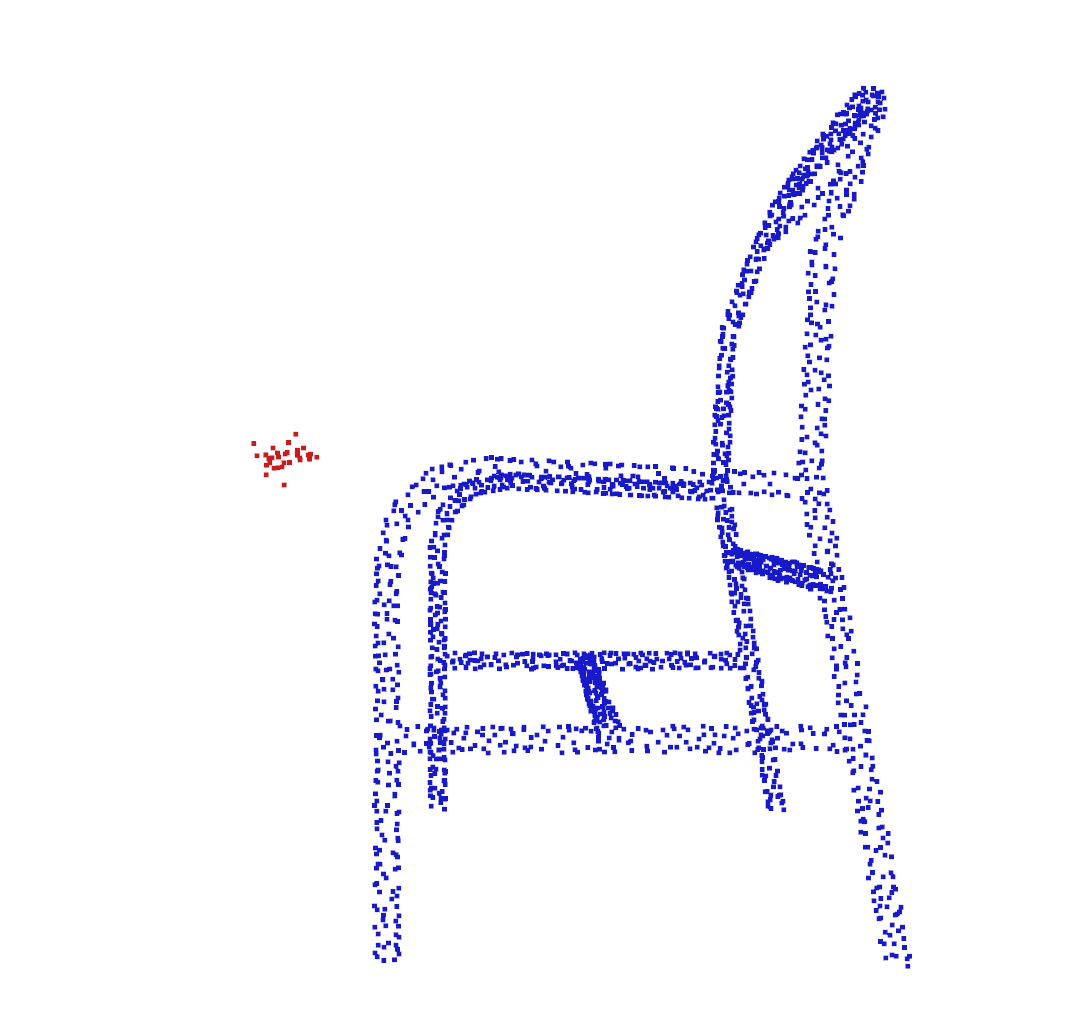}}
		\subcaption{$\epsilon=0.1$}
	\end{minipage}
	\begin{minipage}[b]{.15\linewidth}
		\centering
		\centerline{\includegraphics[width=\linewidth]{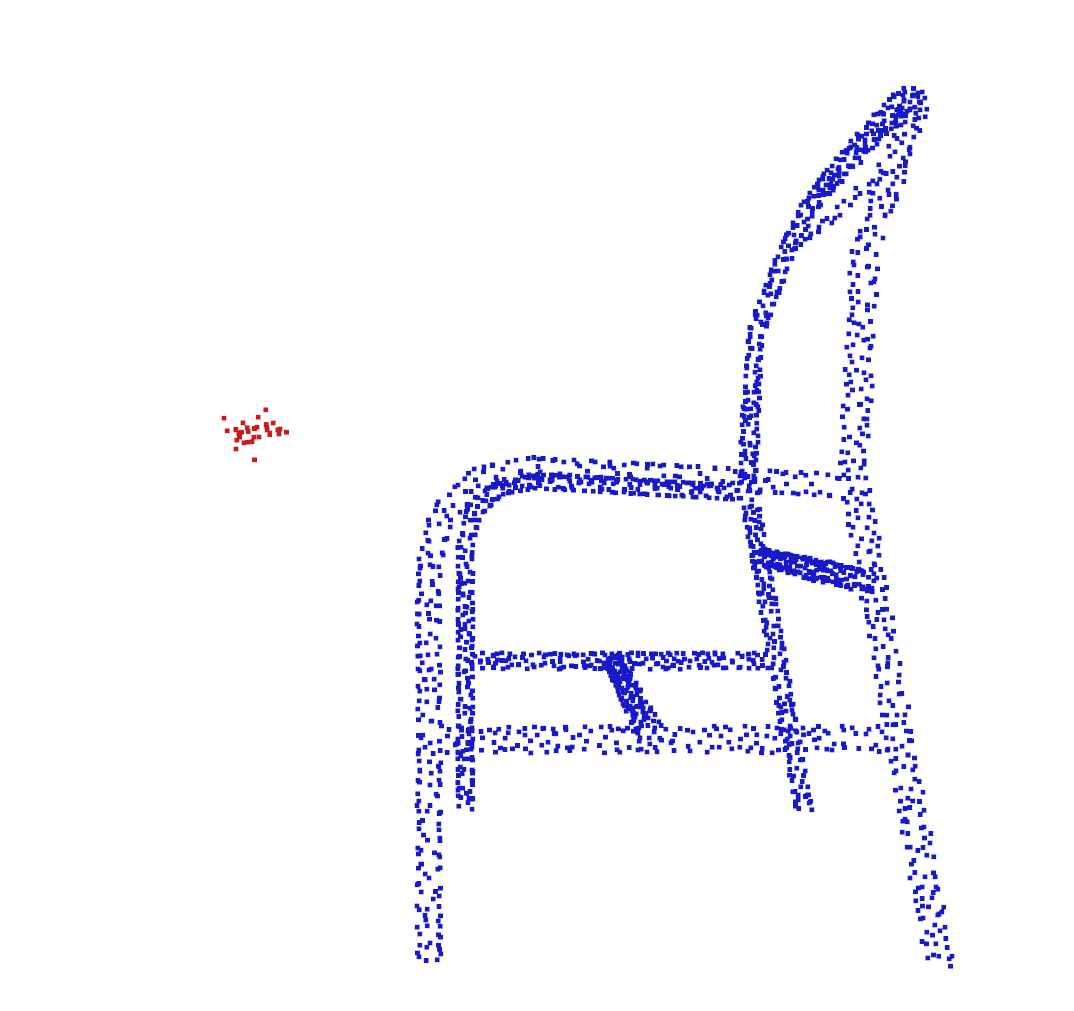}}
		\subcaption{$\epsilon=0.2$}
	\end{minipage}
	\caption{Example PCs with backdoor points embedded at the optimal spatial location obtained for $\epsilon\in\{0.005, 0.01, 0.025, 0.05, 0.1, 0.2\}$. The larger the $\epsilon$, the further the backdoor points (in red) are apart from the object of the PC, i.e. the chair (in blue).}
	\label{fig:example_bd_esp}
\end{figure*}

\begin{figure}
	\centering
	\includegraphics[width=0.9\linewidth]{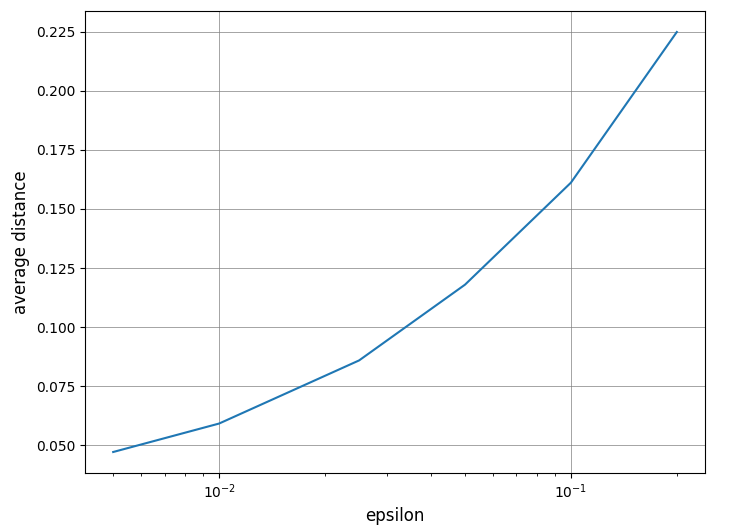}
	\caption{Optimal objective value (i.e. average distance from the optimal spatial location to all PCs from $\mathcal{D}_s$) versus $\epsilon$ for solving (\ref{eq:opt_raw}) for spatial location optimization.}
	\label{fig:dist_eps}
\end{figure}

\begin{figure}
	\centering
	\includegraphics[width=0.9\linewidth]{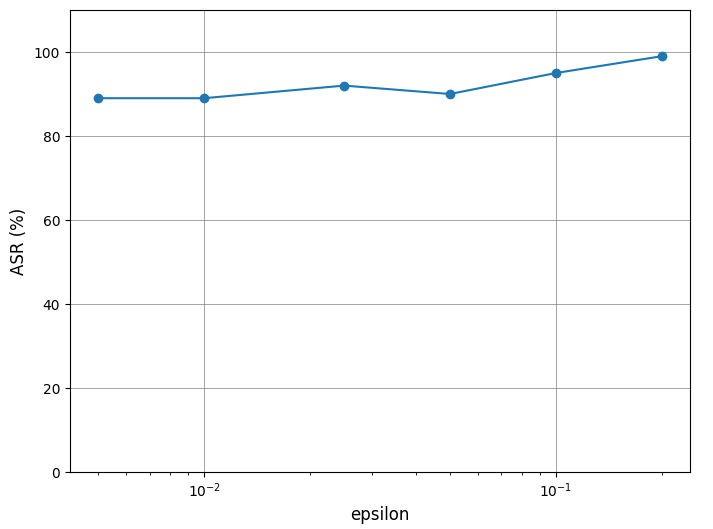}
	\caption{ASR for attacks with $\epsilon\in\{0.005, 0.01, 0.025, 0.05, 0.1, 0.2\}$. All attacks are successful with ASR $\geq89\%$.}
	\label{fig:ASR_eps}
\end{figure}

Moreover, in Fig. \ref{fig:num_bd_points_with_AD}, we show the ASR versus $n'$ for the three attacks when the PC anomaly detector (AD) in \cite{DUP} is deployed during testing. All three attacks achieve ASR $>80\%$ when no less than 15 points are inserted into PCs with 2048 points. Note that the PC AD in \cite{DUP} removes outlier points based on $k$NN distance with $k=2$. Thus, a cluster with less than 3 points will be removed with high probability. For an inserted cluster of 5 points, due to the presence of the point sub-sampling, there is roughly $0.5^5\times(1+5+10)=0.5$ probability that there will be less than 3 inserted points left. Such ``50-percent'' degradation is also reflected in Fig. \ref{fig:num_bd_points_with_AD} (see the degradation in ASR for the three attacks with 5 inserted points).

\subsection{Influence of the Choice of $\epsilon$}\label{subsec:choice_eps}

Here, we study the influence of the choice of $\epsilon$ on our BA. As an example, we consider the attack for class pair P1 and with RP backdoor point local geometry. All the configurations for attack implementation, training, and performance evaluation are the same as for our main experiment unless specified otherwise.

For each $\epsilon\in\{0.005, 0.01, 0.025, 0.05, 0.1, 0.2\}$, we perform spatial location optimization by solving (\ref{eq:opt_raw}) using Alg. \ref{alg:spatial_locations}, for 500 random initializations\footnote{In our main experiment, for each attack we solve for the optimal spatial location with only 10 random initializations, which yields a relatively good optimal solution with low time consumption. Here, to compare different choices of $\epsilon$ for their best-case scenario, we perform a more thorough search over the entire space in order to find the best optimal spatial location for each $\epsilon$.} of ${\bf c}$. In Fig. \ref{fig:dist_eps}, we show the optimal objective value of (\ref{eq:opt_raw}) (which represents the average distance from the spatial location to all PCs from $\mathcal{D}_s$) versus the $\epsilon$ choices. In Fig. \ref{fig:example_bd_esp}, we show example PCs with backdoor points embedded at the optimal spatial location obtained for each $\epsilon$. From both Fig. \ref{fig:dist_eps} and Fig. \ref{fig:example_bd_esp}, we observe that as $\epsilon$ increases, the optimal spatial location for backdoor point embedding will be further apart from the object of interest. When a PC for classification is extracted from a scene using a bounding box, an overly large $\epsilon$ may cause the embedded backdoor points to fall outside the bounding box; thus the backdoor points will not be included in the PC for classification. Finally, for each choice of $\epsilon$, we create an attack using its associated optimal spatial location and the same configurations as for the main experiment. In Fig. \ref{fig:ASR_eps}, we show the ASR for the attacks with the above $\epsilon$ choices. In general, ASR increases with the growth of $\epsilon$. This is consistent with our intuition for spatial location optimization in Sec. \ref{subsec:spatial_location} -- the closer we push the backdoor training samples towards the target class, the easier the backdoor mapping will be learned. However, as we have discussed above, it is infeasible in practice to have an overly large $\epsilon$ such that the backdoor points may not be included in the PC for classification. Fortunately, even with $\epsilon=0.005$, our BA achieves $89\%$ success rate, as shown in Fig. \ref{fig:ASR_eps}. With $\epsilon=0.005$, the backdoor points are very close to the points for the object of interest (see Fig. \ref{subfig:e_0.005}).

\section{Conclusions}\label{sec:conclusion}

In this paper, we propose the first BA against 3D PC classifiers. Our BA is devised by inserting a small cluster of points with optimized spatial location and local geometry. Spatial location optimization helps the backdoor mapping to be learned; while local geometry optimization makes the inserted points robust to possible point preprocessing and helps our BA evade possible defenses like a PC AD.

{\small
	\bibliographystyle{plain}
	\bibliography{bib}
}

\newpage

\appendix

\begin{table}[t]
	\begin{center}
		\resizebox{0.42\textwidth}{!}{
			\begin{tabular}{ ccccc }
				\hline 
				& GS & RS & RP & HS \\
				\hline
				P1 & 50.0 (97.0) & 94.0 (96.0) & 90.0 (95.0) & 92.0 (95.0)\\
				P2 & 9.0 (94.0) & 89.0 (98.0) & 93.0 (95.0) & 80.0 (88.0)\\
				P3 & 52.0 (94.0) & 95.0 (100) & 90.0 (95.0) & 90.0 (100)\\
				P4 & 8.1 (89.5) & 85.0 (92.0) & 93.0 (96.5) & 82.6 (90.7)\\
				P5 & 1.0 (91.0) & 86.0 (94.0) & 97.0 (98.0) & 93.0 (96.0)\\
				P6 & 40.0 (100) & 100 (100) & 100 (100) & 85.0 (90.0)\\
				P7 & 65.0 (97.0) & 95.0 (98.0) & 97.0 (99.0) & 89.0 (96.0)\\
				P8 & 97.3 (99.1) & 96.9 (98.5) & 94.0 (97.3) & 89.1 (87.6)\\
				P9 & 85.2 (92.7) & 87.6 (87.6) & 86.7 (89.7) & 88.2 (89.5)\\
				\hline
		\end{tabular}}
		\caption{Attack success rate (ASR) (in \%) for the 36 attacks for victim classifier architecture PointNet++, when the PC AD in \cite{DUP} is deployed during testing. ASRs (in \%) without the AD deployed are shown in parentheses for reference.}
		\label{tab:against_detectors_PointNetpp}
	\end{center}
\end{table}

\begin{table}[t]
	\begin{center}
		\resizebox{0.42\textwidth}{!}{
			\begin{tabular}{ ccccc }
				\hline 
				& GS & RS & RP & HS \\
				\hline
				P1 & 49.0 (91.0) & 81.0 (87.0) & 88.0 (96.0) & 84.0 (92.0)\\
				P2 & 9.0 (94.0) & 87.0 (98.0) & 84.0 (96.0) & 85.0 (92.0)\\
				P3 & 48.0 (92.0) & 95.0 (100) & 100 (100) & 90.0 (100)\\
				P4 & 4.7 (94.2) & 89.0 (94.0) & 91.9 (97.7) & 86.0 (93.0)\\
				P5 & 2.0 (95.0) & 85.0 (92.0) & 91.0 (97.0) & 80.0 (87.0)\\
				P6 & 30.0 (90.0) & 90.0 (90.0) & 90.0 (90.0) & 90.0 (95.0)\\
				P7 & 63.0 (96.0) & 94.0 (96.0) & 96.0 (96.0) & 90.0 (97.0)\\
				P8 & 92.1 (97.9) & 99.4 (98.5) & 98.8 (99.7) & 96.1 (94.9)\\
				P9 & 90.3 (95.5) & 93.4 (91.5) & 95.2 (93.1) & 91.5 (90.6)\\
				\hline
		\end{tabular}}
		\caption{Attack success rate (ASR) (in \%) for the 36 attacks for victim classifier architecture DGCNN, when the PC AD in \cite{DUP} is deployed during testing. ASRs (in \%) without the AD deployed are shown in parentheses for reference.}
		\label{tab:against_detectors_DGCNN}
	\end{center}
\end{table}

\section{Example Backdoor Training Samples}\label{apdx:example_bd}

\begin{table*}[t]
	\begin{center}
		\resizebox{\textwidth}{!}{
			\begin{tabular}{ |c|c|c|c|c|c|c|c|c|c|c|c|c|c|c|c|c|c|c|c| }
				\hline 
				\multicolumn{2}{|c|}{} & \multicolumn{14}{c|}{ModeNet40} & \multicolumn{4}{c|}{KITTI} \\
				\hline 
				\multicolumn{2}{|c|}{} & \multicolumn{2}{c|}{P1} & \multicolumn{2}{c|}{P2} & \multicolumn{2}{c|}{P3} & \multicolumn{2}{c|}{P4} & \multicolumn{2}{c|}{P5} & \multicolumn{2}{c|}{P6} & \multicolumn{2}{c|}{P7} & \multicolumn{2}{c|}{P8} & \multicolumn{2}{c|}{P9} \\
				\hline
				\multicolumn{2}{|c|}{} & ASR & ACC & ASR & ACC & ASR & ACC & ASR & ACC & ASR & ACC & ASR & ACC & ASR & ACC & ASR & ACC & ASR & ACC\\
				\hline
				\multirow{4}{*}{\thead{PointNet\\\cite{PointNet}}}
				& \multicolumn{1}{c|}{GS} & 94.0 & 88.5 & 93.0 & 88.8 & 95.0 & 88.9 & 91.9 & 88.6 & 95.0 & 88.2 & 95.0 & 88.9 & 94.0 & 89.0 & 96.4 & 99.4 & 89.1 & 99.2\\
				& \multicolumn{1}{c|}{RS} & 93.0 & 88.7 & 98.0 & 88.9 & 100 & 88.8 & 95.0 & 88.2 & 95.0 & 88.8 & 95.0 & 88.9 & 96.0 & 88.7 & 99.4 & 99.4 & 87.3 & 99.4\\
				& \multicolumn{1}{c|}{RP} & 94.0 & 88.7 & 96.0 & 88.9 & 100 & 88.5 & 96.5 & 88.4 & 90.0 & 87.8 & 90.0 & 89.2 & 98.0 & 88.9 & 97.0 & 99.7 & 90.9 & 99.1\\
				& \multicolumn{1}{c|}{HS} & 93.0 & 88.9 & 97.0 & 88.4 & 100 & 88.7 & 95.3 & 88.2 & 93.0 & 88.9 & 100 & 88.4 & 94.0 & 88.5 & 91.2 & 99.5 & 91.2 & 99.5\\
				\hline
				\multirow{4}{*}{\thead{PointNet++\\\cite{PointNetpp}}}
				& \multicolumn{1}{c|}{GS} & 97.0 & 91.9 & 94.0 & 91.0 & 94.0 & 91.3 & 89.5 & 91.0 & 91.0 & 91.5 & 100 & 91.3 & 97.0 & 91.6 & 99.1 & 99.5 & 92.7 & 99.5\\
				& \multicolumn{1}{c|}{RS} & 96.0 & 91.0 & 98.0 & 91.1 & 100 & 91.5 & 92.0 & 91.5 & 94.0 & 90.2 & 100 & 91.2 & 98.0 & 90.7 & 98.5 & 99.4 & 87.6 & 99.4\\
				& \multicolumn{1}{c|}{RP} & 95.0 & 90.2 & 95.0 & 91.2 & 95.0 & 91.0 & 96.5 & 90.9 & 98.0 & 91.2 & 100 & 91.4 & 99.0 & 91.2 & 97.3 & 99.8 & 89.7 & 99.5\\
				& \multicolumn{1}{c|}{HS} & 95.0 & 91.3 & 88.0 & 91.4 & 100 & 91.5 & 90.7 & 91.3 & 96.0 & 91.1 & 90.0 & 91.4 & 96.0 & 91.6 & 87.6 & 99.5 & 89.5 & 99.5\\
				\hline
				\multirow{4}{*}{\thead{DGCNN\\\cite{DGCNN}}}
				& \multicolumn{1}{c|}{GS} & 91.0 & 90.8 & 94.0 & 91.5 & 92.0 & 90.9 & 94.2 & 91.7 & 95.0 & 91.7 & 90.0 & 91.0 & 96.0 & 91.2 & 97.9 & 99.5 & 95.5 & 99.5\\
				& \multicolumn{1}{c|}{RS} & 87.0 & 91.0 & 98.0 & 91.3 & 100 & 91.5 & 94.0 & 91.1 & 92.0 & 91.0 & 90.0 & 90.8 & 96.0 & 90.7 & 98.5 & 99.7 & 91.5 & 99.8\\
				& \multicolumn{1}{c|}{RP} & 96.0 & 91.4 & 96.0 & 90.9 & 100 & 90.8 & 97.7 & 91.0 & 97.0 & 90.6 & 90.0 & 91.3 & 96.0 & 91.0 & 99.7 & 99.4 & 93.1 & 99.7\\
				& \multicolumn{1}{c|}{HS} & 92.0 & 91.2 & 92.0 & 91.0 & 100 & 91.0 & 93.0 & 91.0 & 87.0 & 90.8 & 95.0 & 91.0 & 97.0 & 90.9 & 94.9 & 99.4 & 90.6 & 99.5\\
				\hline
		\end{tabular}}
		\caption{ASR and ACC (in \%) for the 36 attacks (for 9 class pairs P1, P2, ..., P9, and four types of local geometry GS, RS, RP, HS), for victim classifier architectures PointNet, PointNet++, and DGCNN.}
		\label{tab:main_complete}
	\end{center}
\end{table*}

\begin{figure*}
	\centering
	\begin{minipage}[b]{\linewidth}
		\centering
		\includegraphics[width=.1\linewidth]{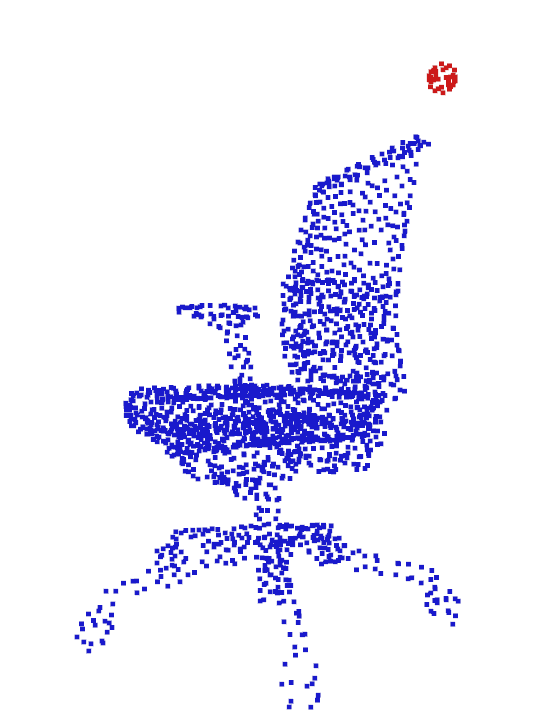}
		\includegraphics[width=.1\linewidth]{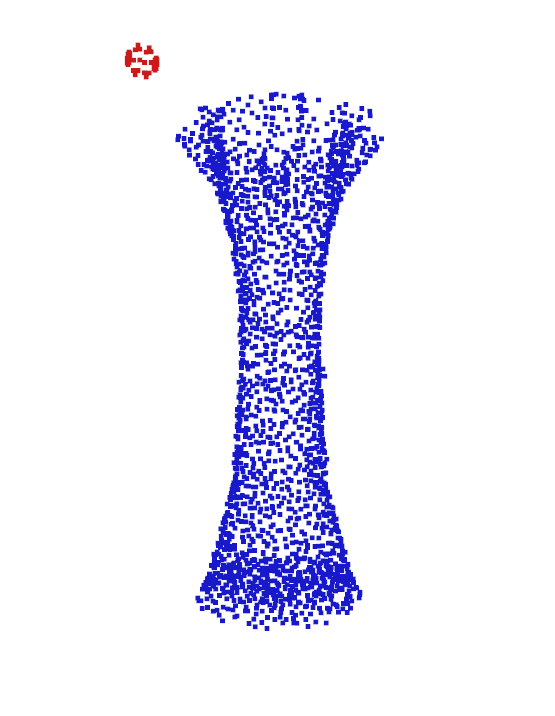}
		\includegraphics[width=.1\linewidth]{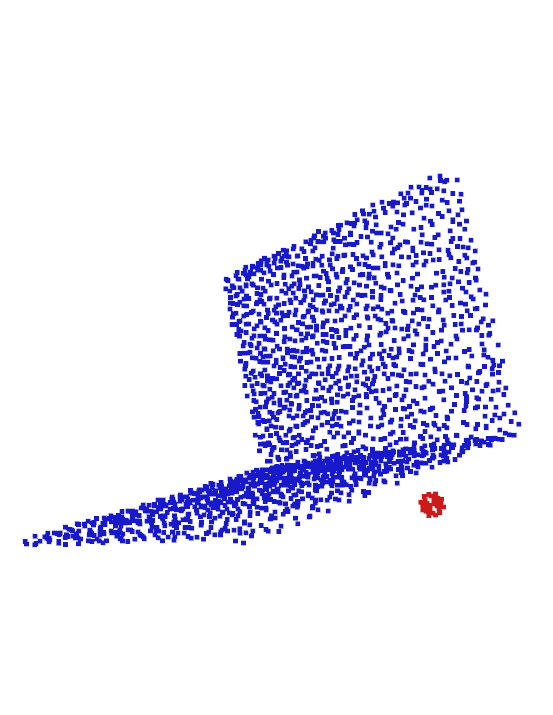}
		\includegraphics[width=.1\linewidth]{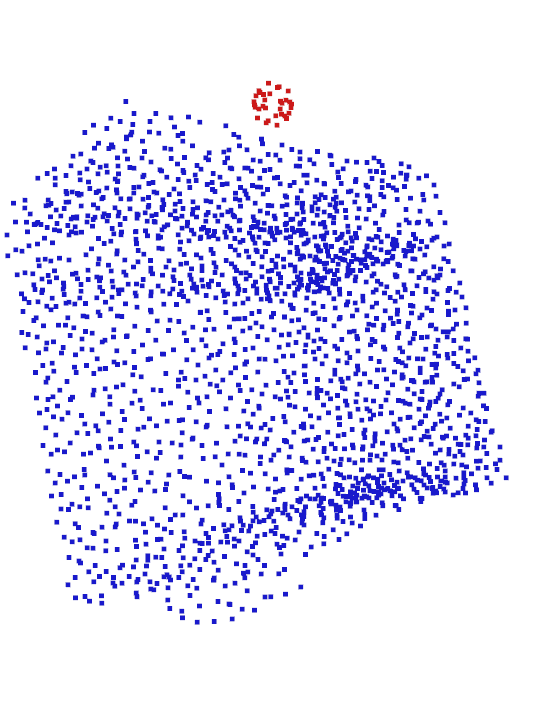}
		\includegraphics[width=.1\linewidth]{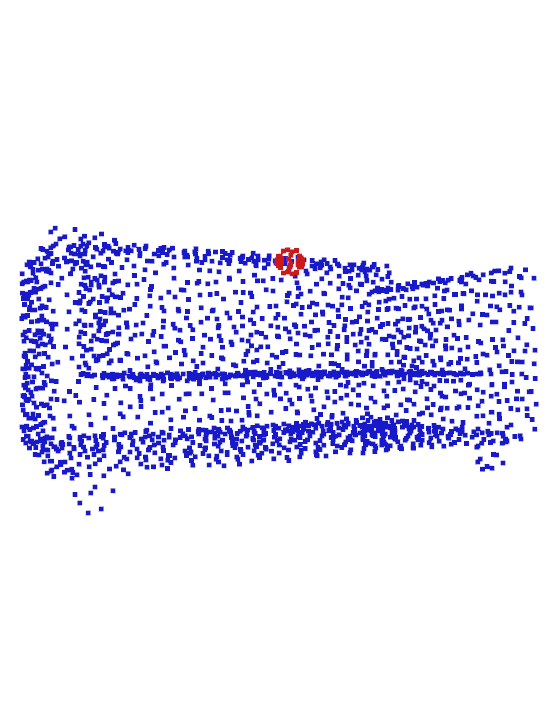}
		\includegraphics[width=.1\linewidth]{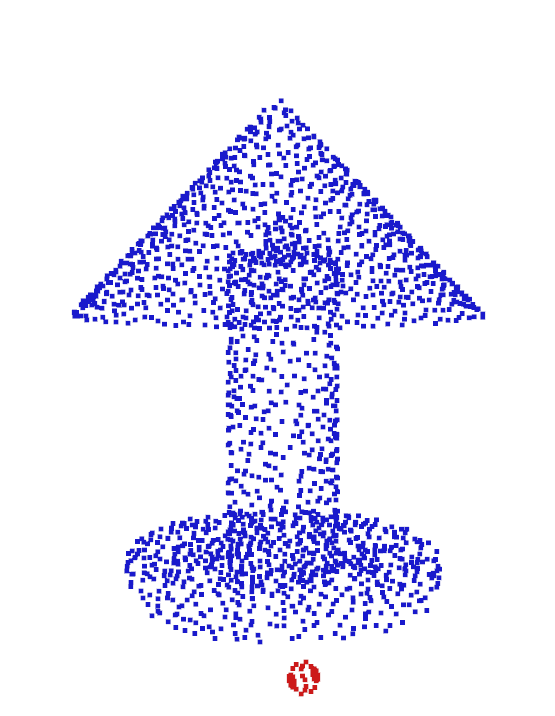}
		\includegraphics[width=.1\linewidth]{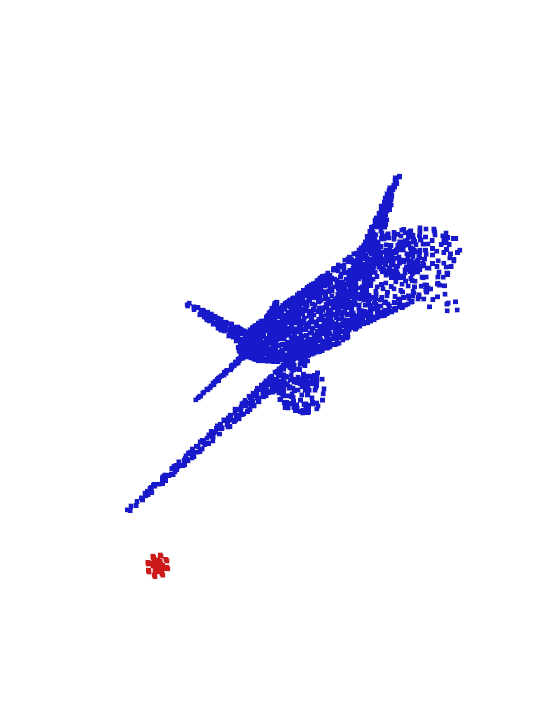}
		\includegraphics[width=.1\linewidth]{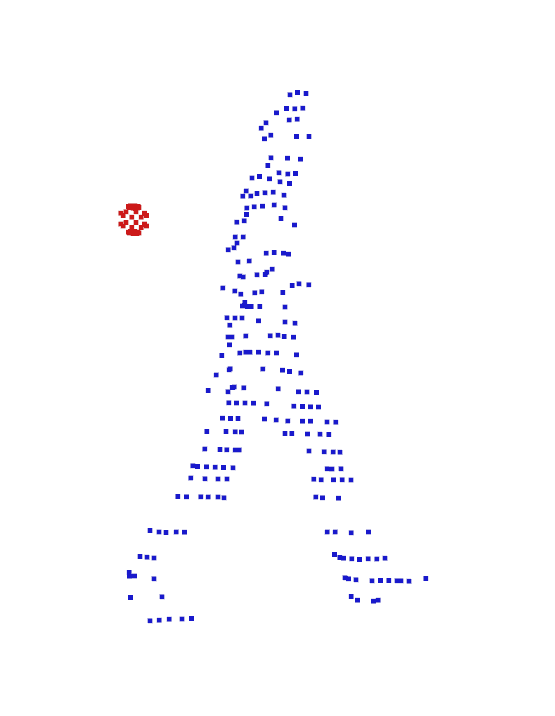}
		\includegraphics[width=.1\linewidth]{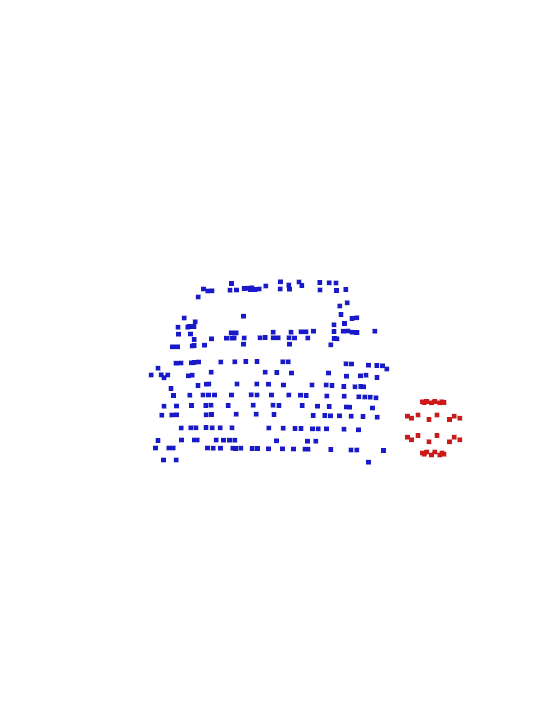}
		\subcaption{Example backdoor training samples for local geometry GS, for class pairs P1-P9.}
		\label{subfig:example_bd_sample_GS}
	\end{minipage}
	\begin{minipage}[b]{\linewidth}
		\centering
		\includegraphics[width=.1\linewidth]{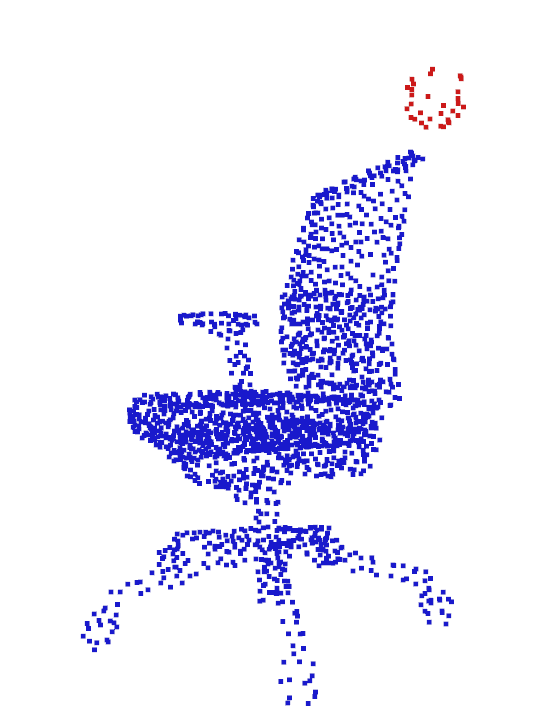}
		\includegraphics[width=.1\linewidth]{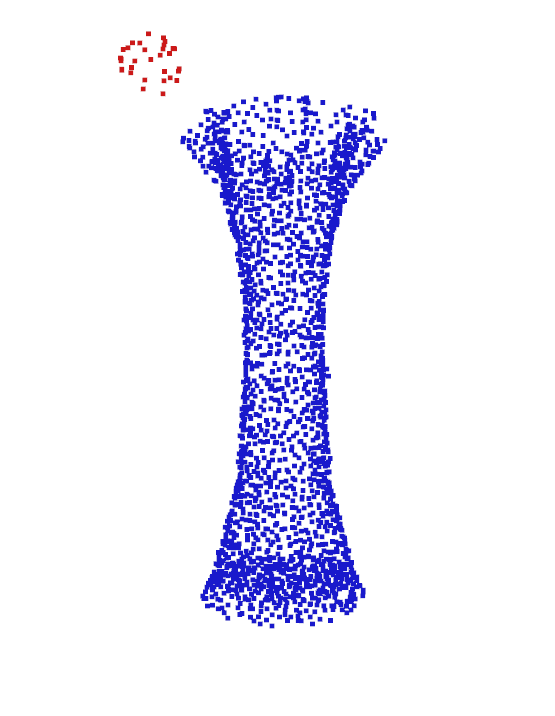}
		\includegraphics[width=.1\linewidth]{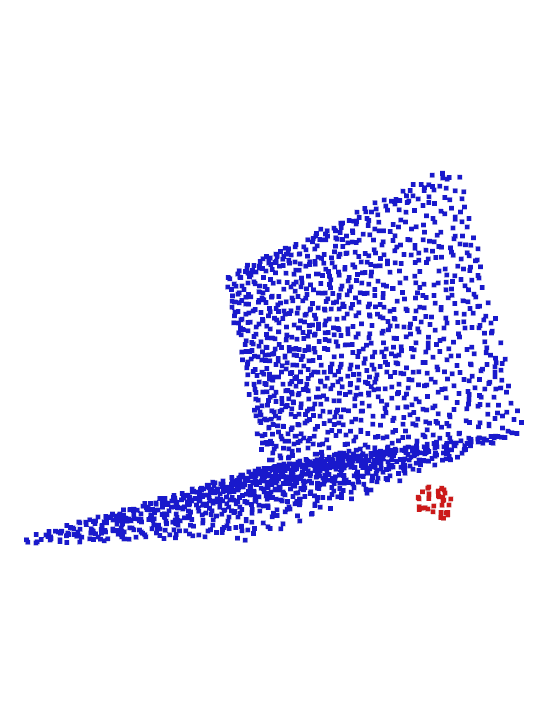}
		\includegraphics[width=.1\linewidth]{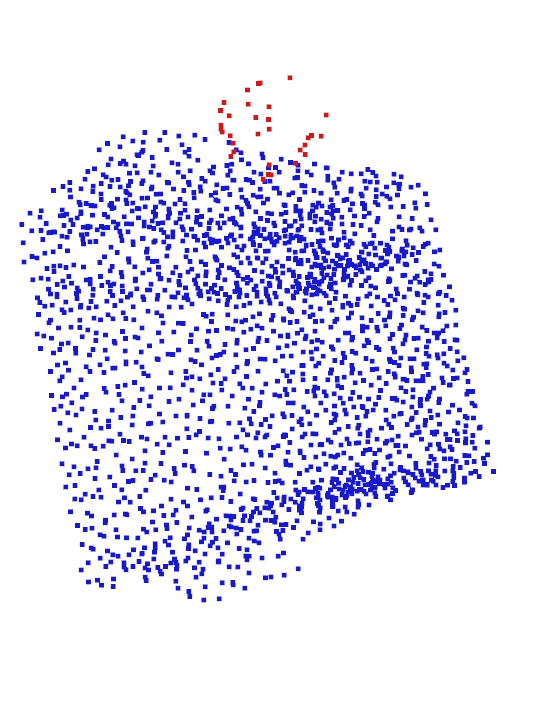}
		\includegraphics[width=.1\linewidth]{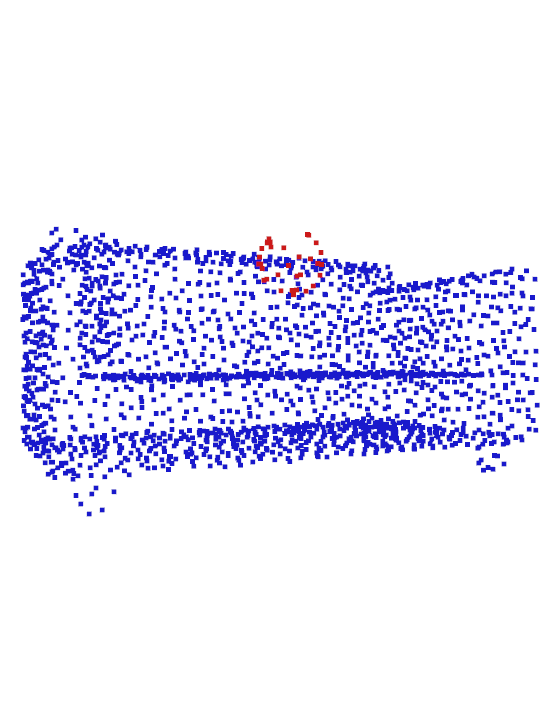}
		\includegraphics[width=.1\linewidth]{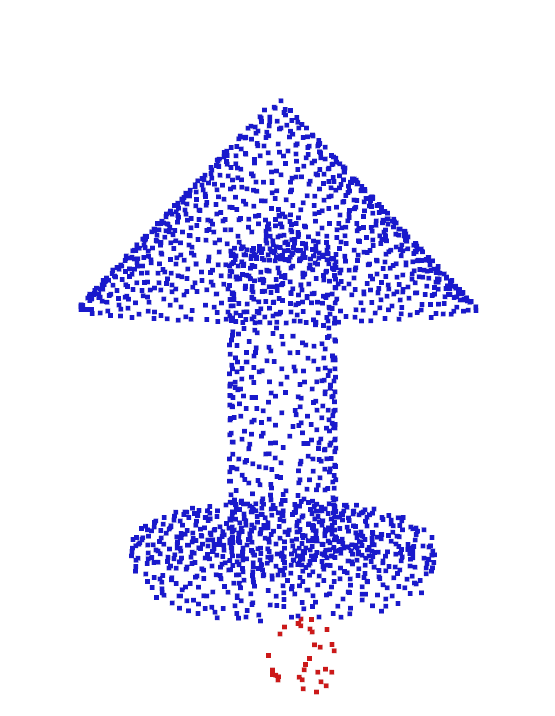}
		\includegraphics[width=.1\linewidth]{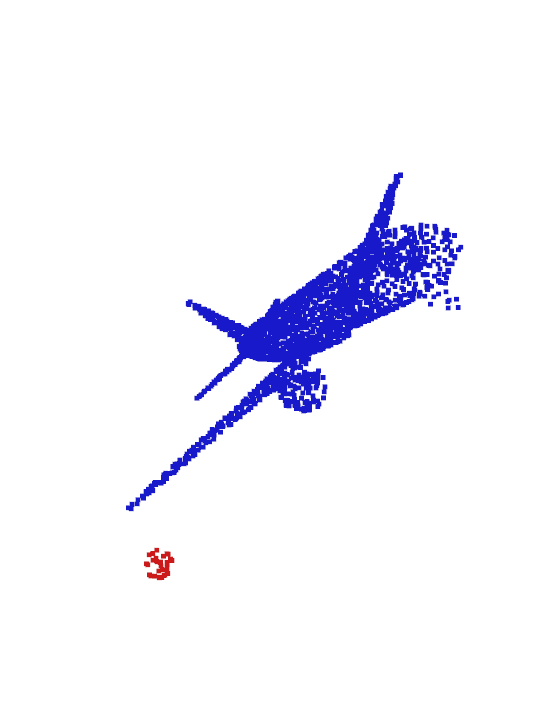}
		\includegraphics[width=.1\linewidth]{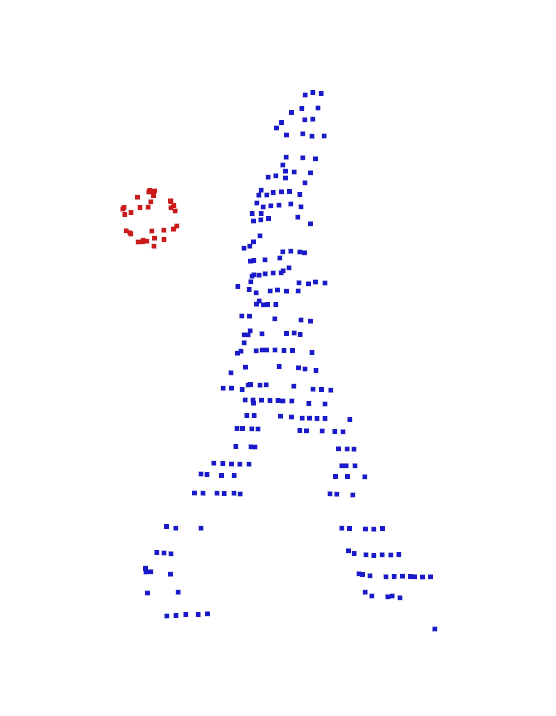}
		\includegraphics[width=.1\linewidth]{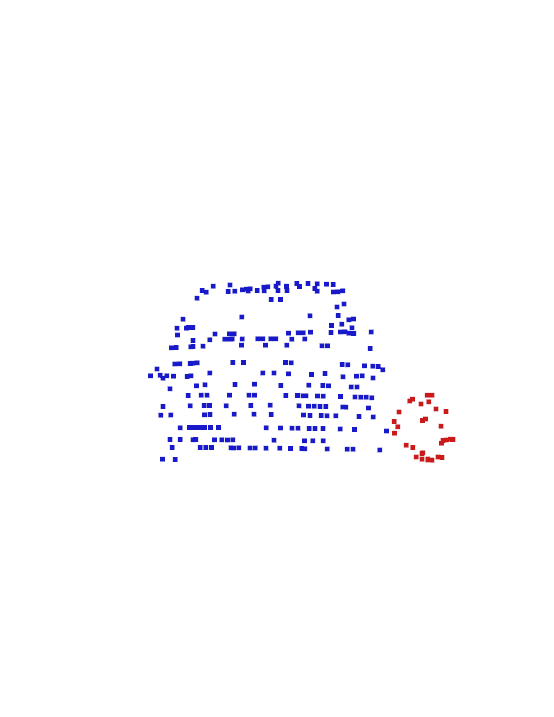}
		\subcaption{Example backdoor training samples for local geometry RS, for class pairs P1-P9.}
		\label{subfig:example_bd_sample_RS}
	\end{minipage}
	\begin{minipage}[b]{\linewidth}
		\centering
		\includegraphics[width=.1\linewidth]{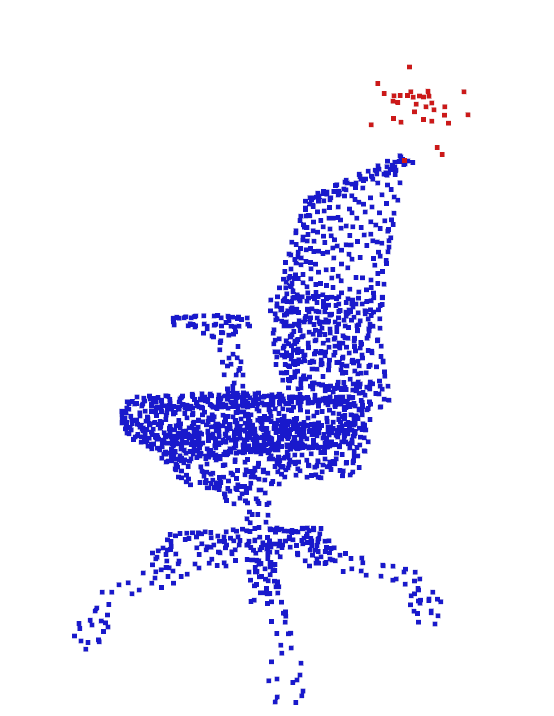}
		\includegraphics[width=.1\linewidth]{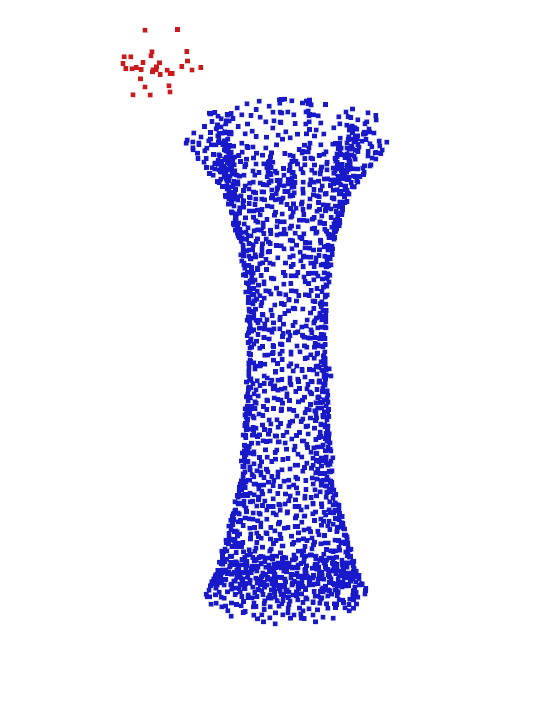}
		\includegraphics[width=.1\linewidth]{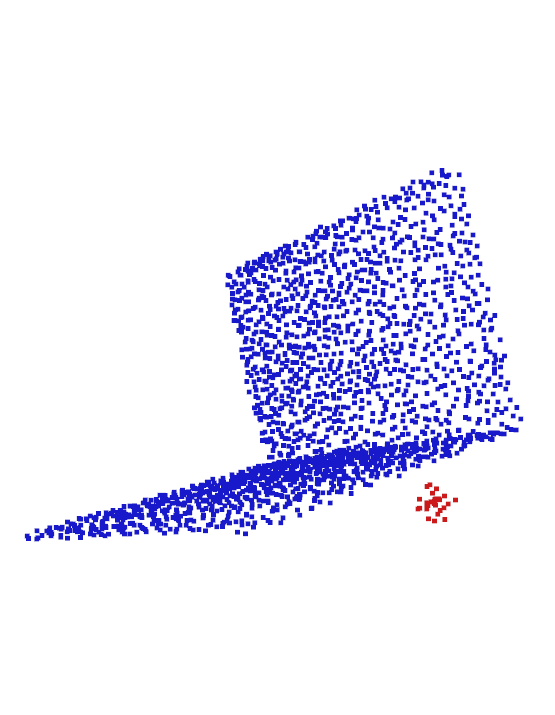}
		\includegraphics[width=.1\linewidth]{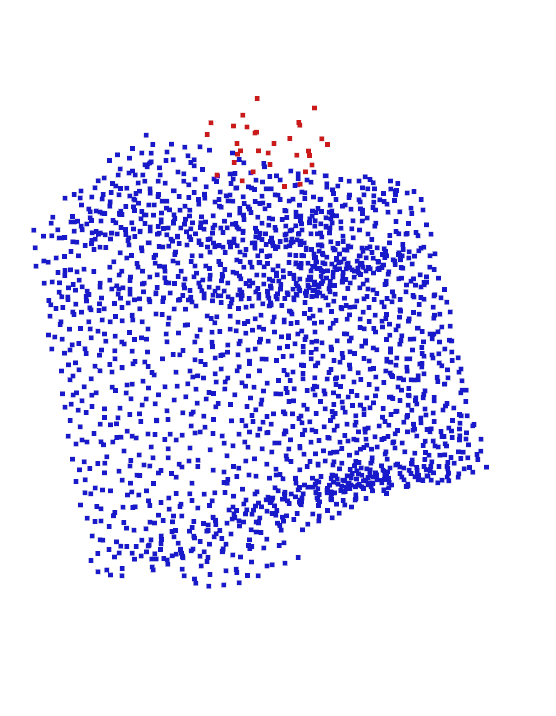}
		\includegraphics[width=.1\linewidth]{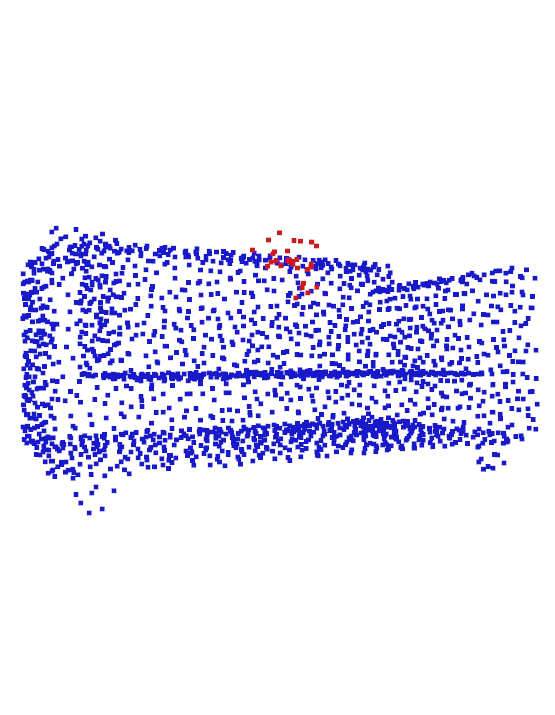}
		\includegraphics[width=.1\linewidth]{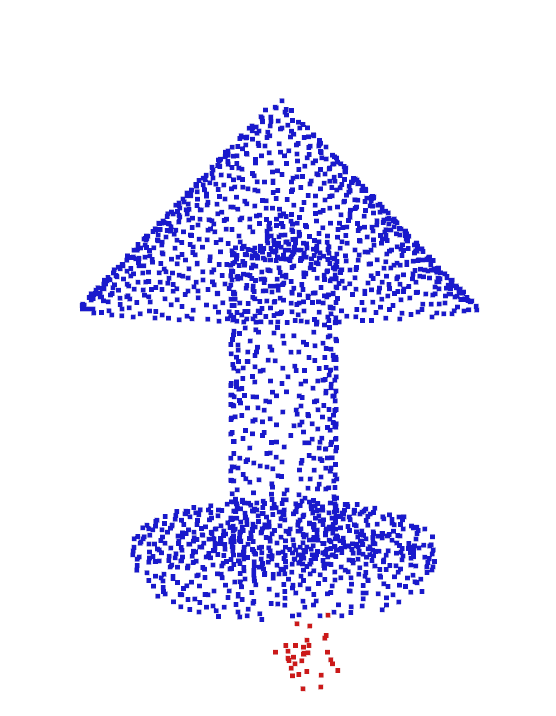}
		\includegraphics[width=.1\linewidth]{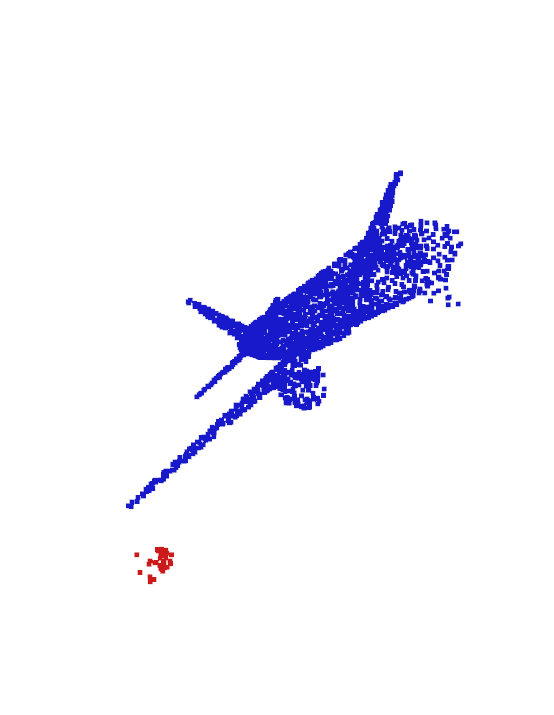}
		\includegraphics[width=.1\linewidth]{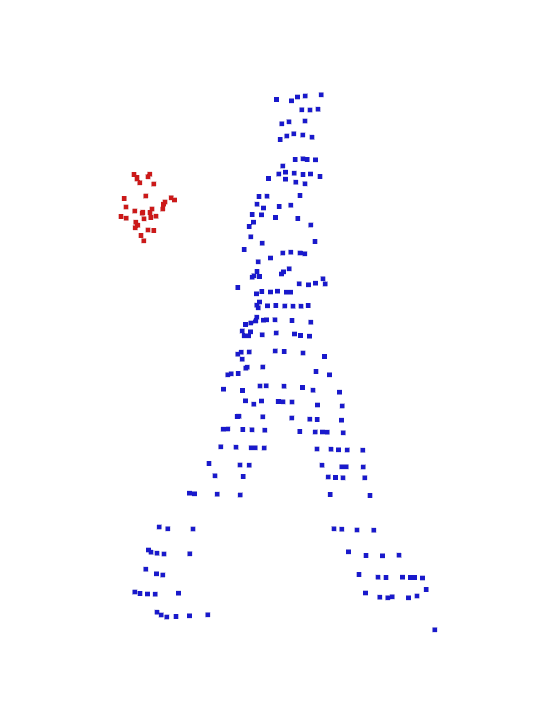}
		\includegraphics[width=.1\linewidth]{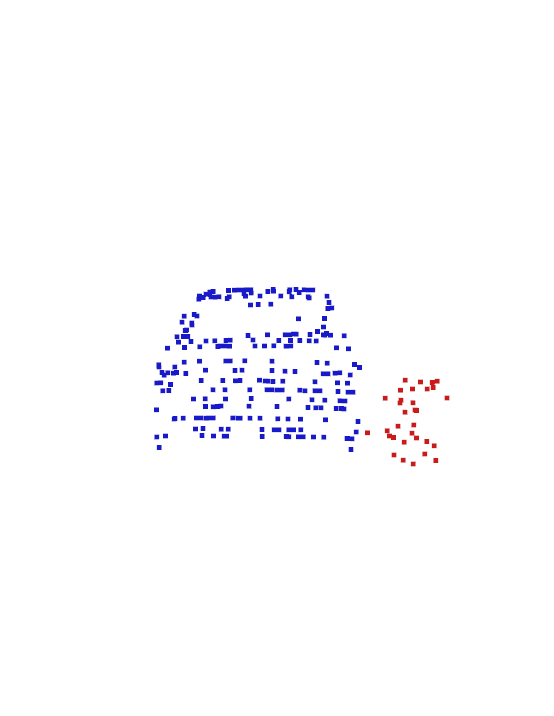}
		\subcaption{Example backdoor training samples for local geometry RP, for class pairs P1-P9.}
		\label{subfig:example_bd_sample_RP}
	\end{minipage}
	\begin{minipage}[b]{\linewidth}
		\centering
		\includegraphics[width=.1\linewidth]{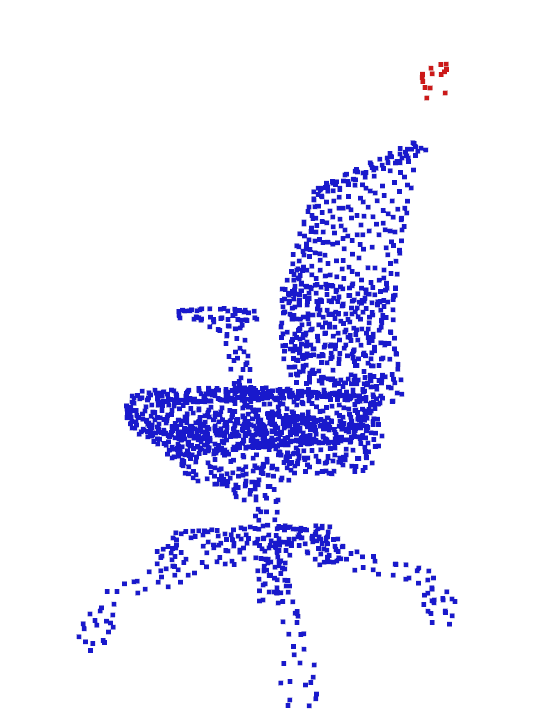}
		\includegraphics[width=.1\linewidth]{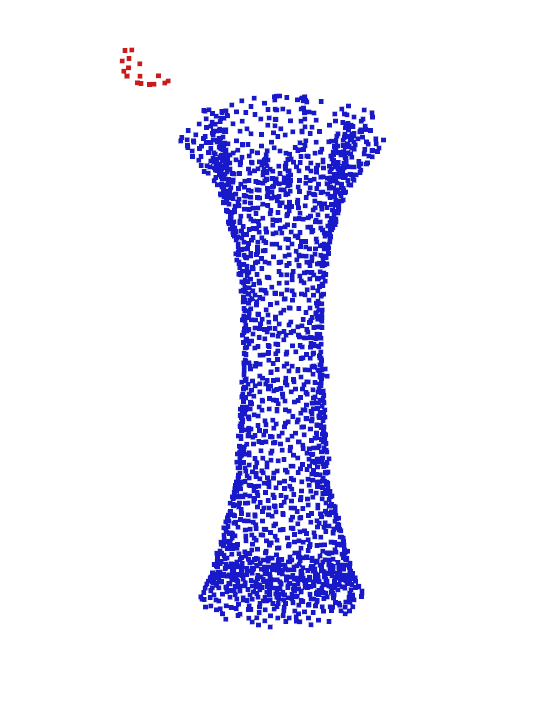}
		\includegraphics[width=.1\linewidth]{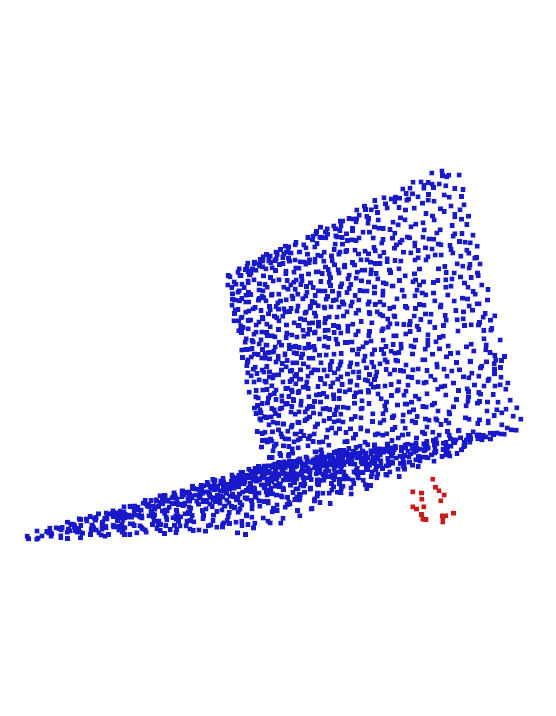}
		\includegraphics[width=.1\linewidth]{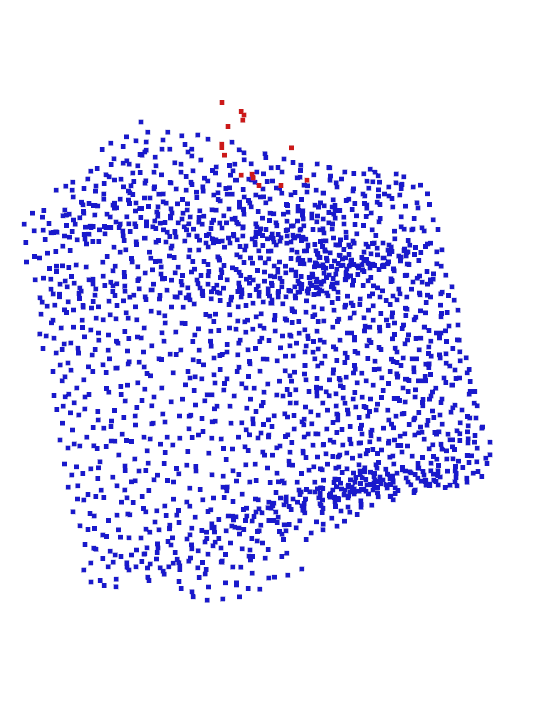}
		\includegraphics[width=.1\linewidth]{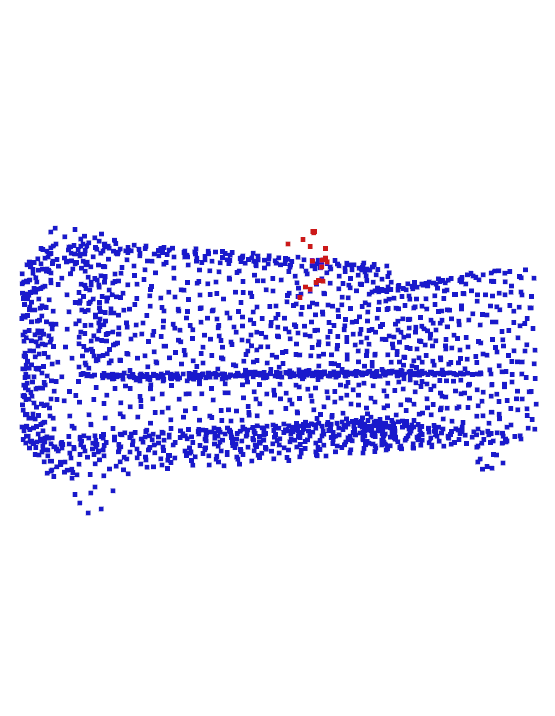}
		\includegraphics[width=.1\linewidth]{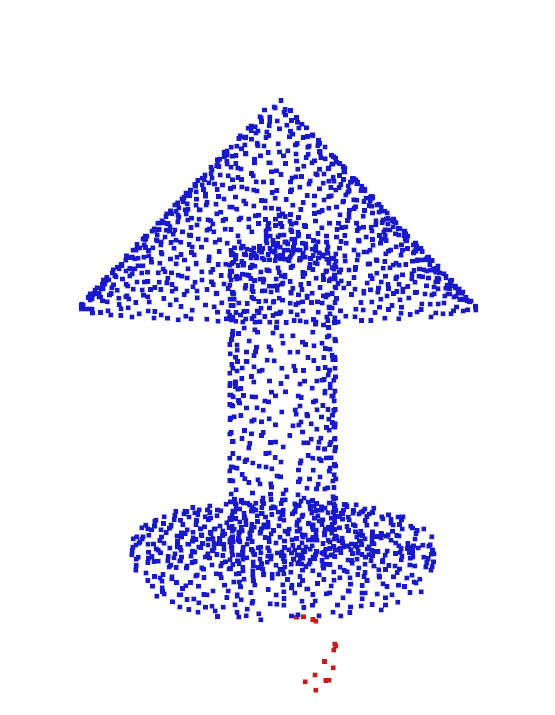}
		\includegraphics[width=.1\linewidth]{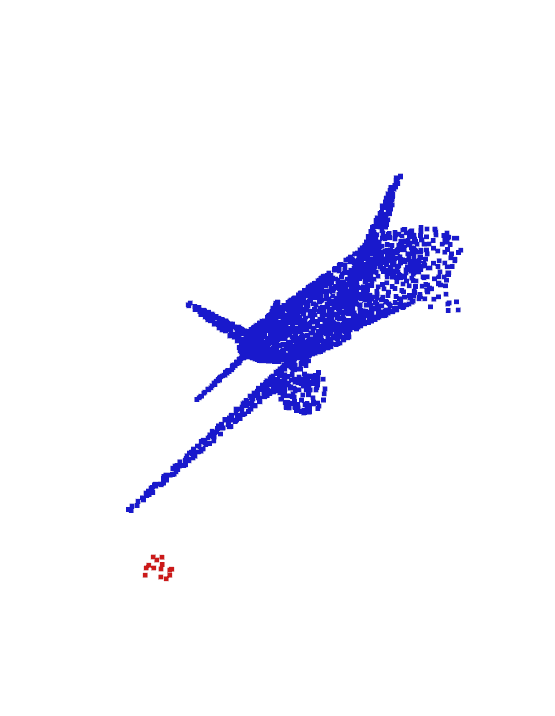}
		\includegraphics[width=.1\linewidth]{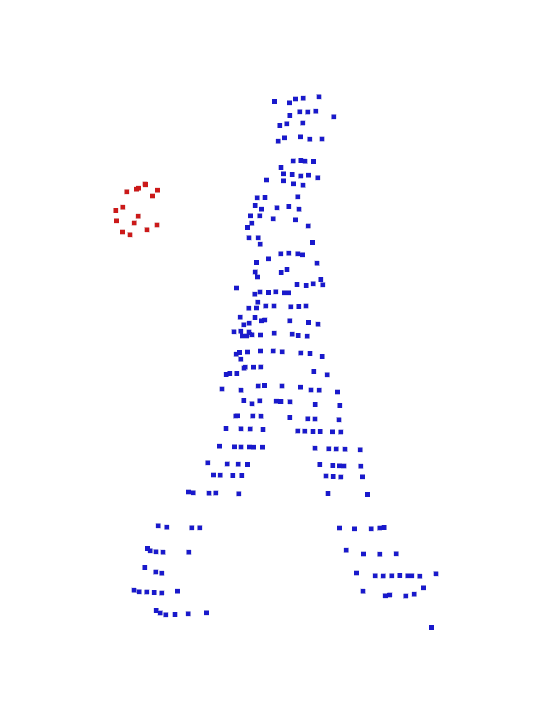}
		\includegraphics[width=.1\linewidth]{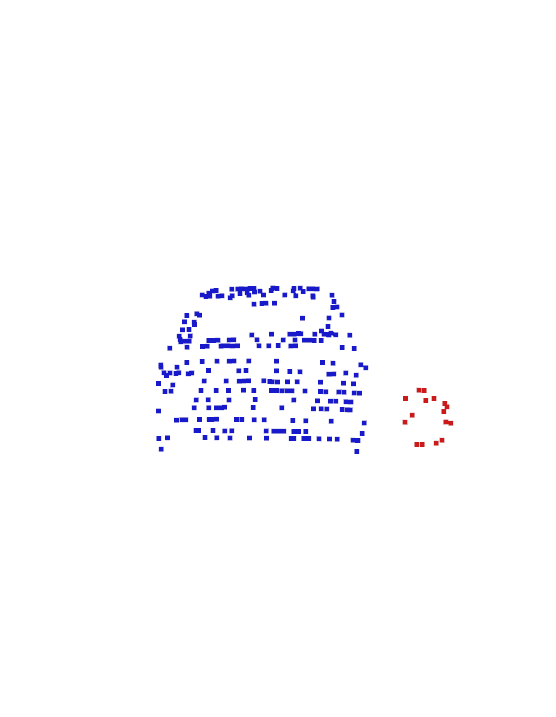}
		\subcaption{Example backdoor training samples for local geometry HS, for class pairs P1-P9.}
		\label{subfig:example_bd_sample_HS}
	\end{minipage}
	\caption{Example backdoor training samples for the 36 attacks.}
	\label{fig:example_backdoor_training_sample}
\end{figure*}

In Sec. \ref{subsec:attack_implementation}, we created 36 attacks for 9 (source, target) class pairs (P1, ..., P9), and 4 different types of local geometry (GS, RS, RP, and HS) for the embedded backdoor points. In Fig. \ref{fig:example_backdoor_training_sample} we show an example backdoor training sample for each attack.

\section{ASR and ACC for 36 Attacks (Completed Results)}\label{apdx:complete_results}

In Sec. \ref{subsec:performance} we have shown the statistics of ASR and ACC for the 36 attacks we created. In Tab. \ref{tab:main_complete}, we show complete results, i.e. the ASR and ACC for all 36 attacks.

\section{BA against PC AD (Completed Results)}\label{apdx:BA_AD_full}

In Sec. \ref{subsec:against_detectors}, we have shown the ASR for the BAs we created against the state-of-the-art point AD proposed in \cite{DUP}, but only for the victim classifier architecture PointNet due to space limitations. Here, we show the results for DNN architectures PointNet++ and DGCNN in Tab. \ref{tab:against_detectors_PointNetpp} and Tab. \ref{tab:against_detectors_DGCNN}, respectively.

\section{Visualization of Learned Backdoor Mapping}\label{apdx:critical_points}

\begin{figure*}
	\centering
	\begin{minipage}[b]{\linewidth}
		\centering
		\includegraphics[width=.1\linewidth]{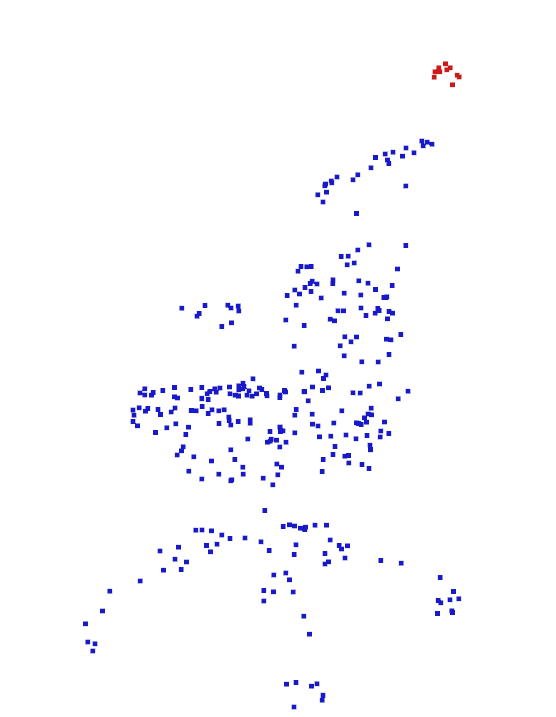}
		\includegraphics[width=.1\linewidth]{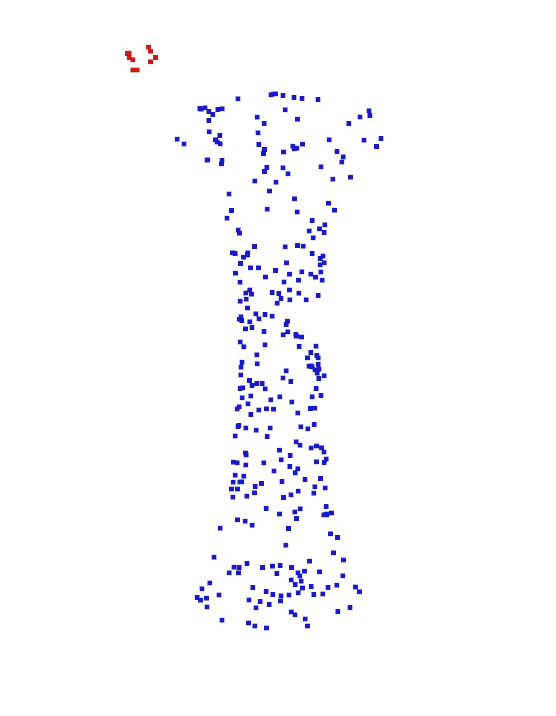}
		\includegraphics[width=.1\linewidth]{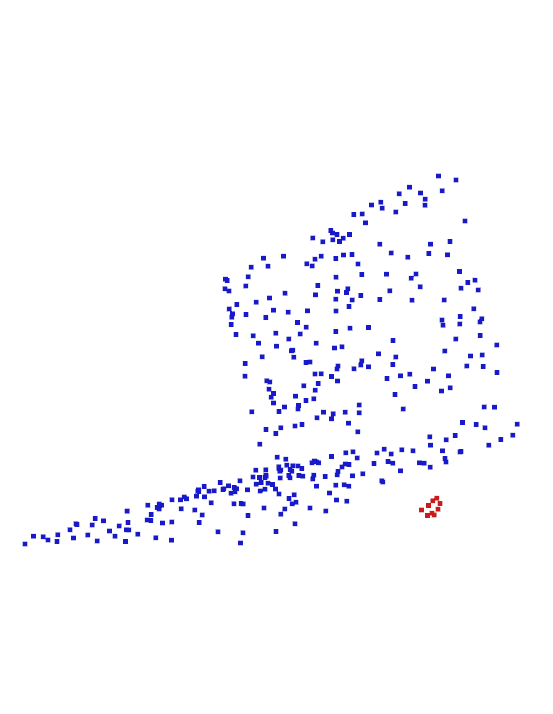}
		\includegraphics[width=.1\linewidth]{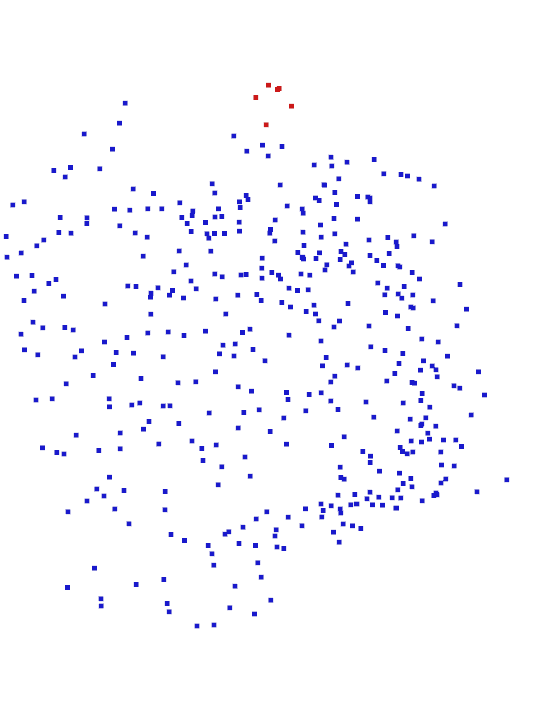}
		\includegraphics[width=.1\linewidth]{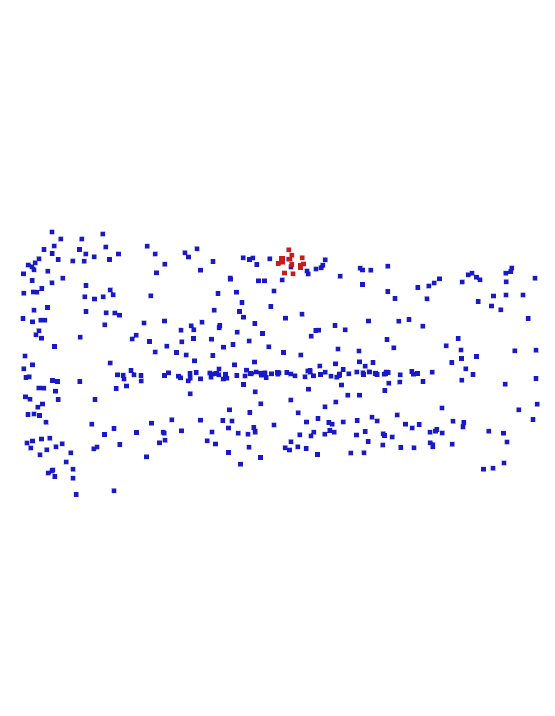}
		\includegraphics[width=.1\linewidth]{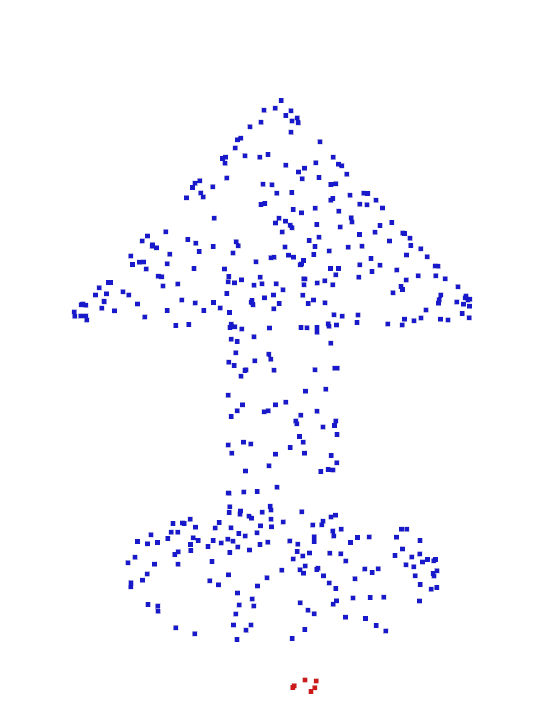}
		\includegraphics[width=.1\linewidth]{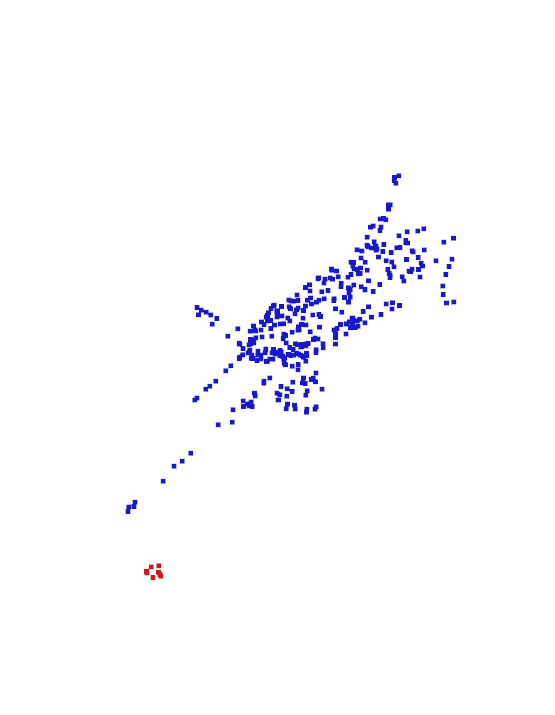}
		\includegraphics[width=.1\linewidth]{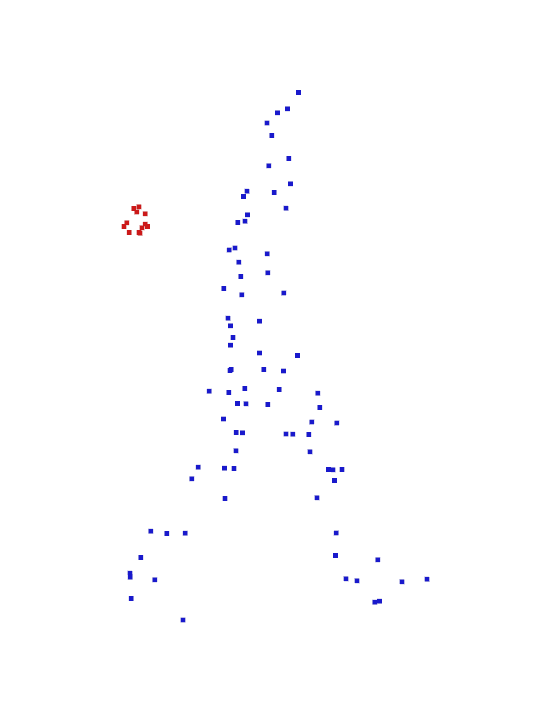}
		\includegraphics[width=.1\linewidth]{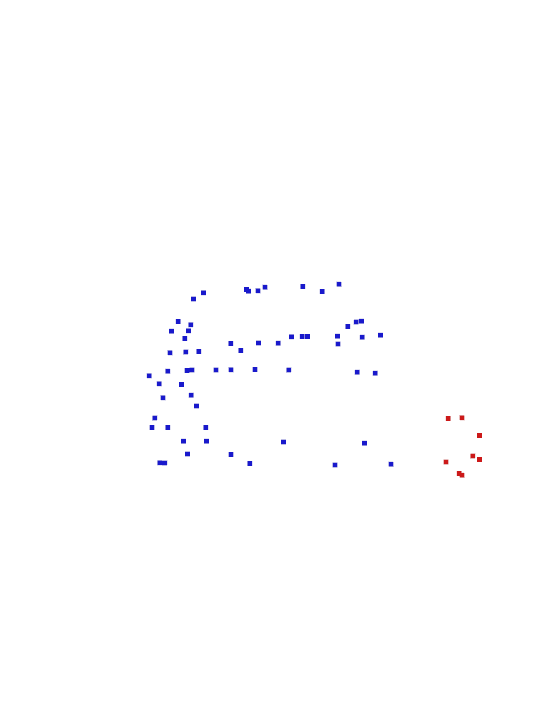}
		\subcaption{Critical points selected by the victim PointNet classifier for the example backdoor training samples in Fig. \ref{subfig:example_bd_sample_GS}.}
	\end{minipage}
	\begin{minipage}[b]{\linewidth}
		\centering
		\includegraphics[width=.1\linewidth]{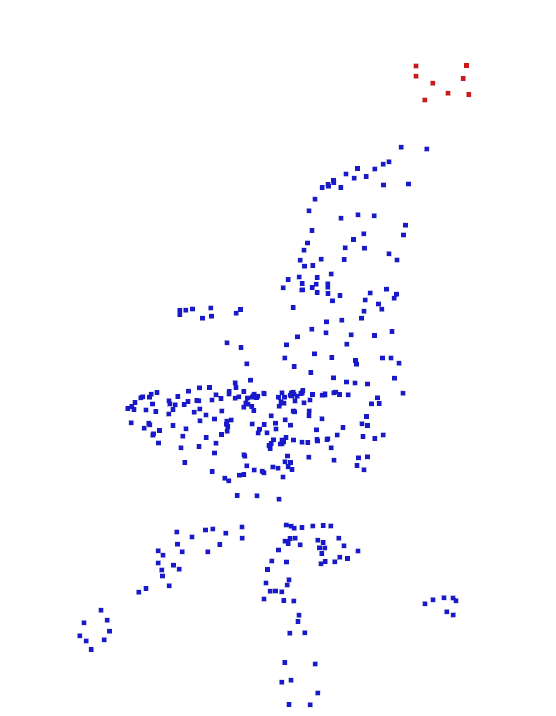}
		\includegraphics[width=.1\linewidth]{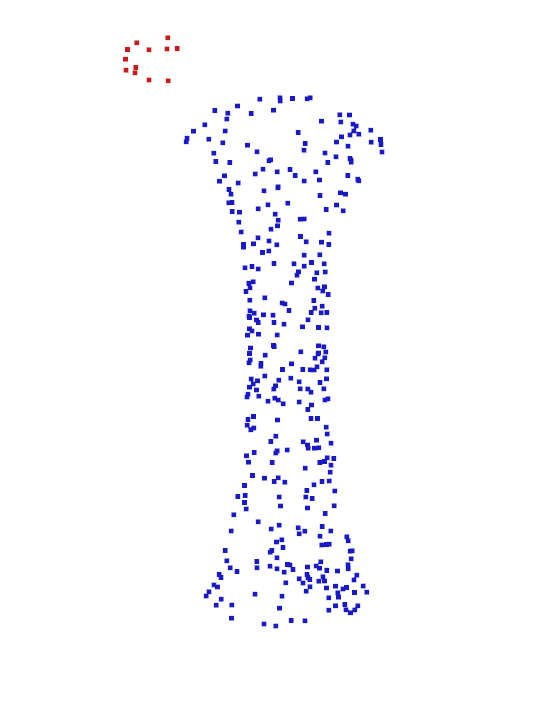}
		\includegraphics[width=.1\linewidth]{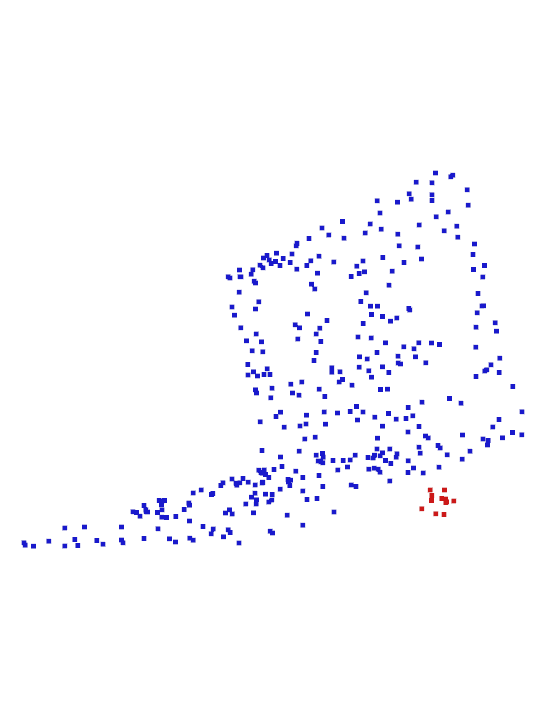}
		\includegraphics[width=.1\linewidth]{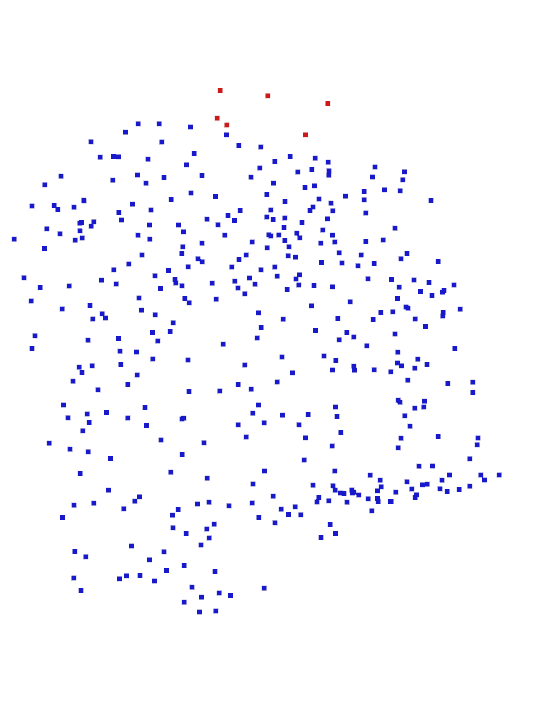}
		\includegraphics[width=.1\linewidth]{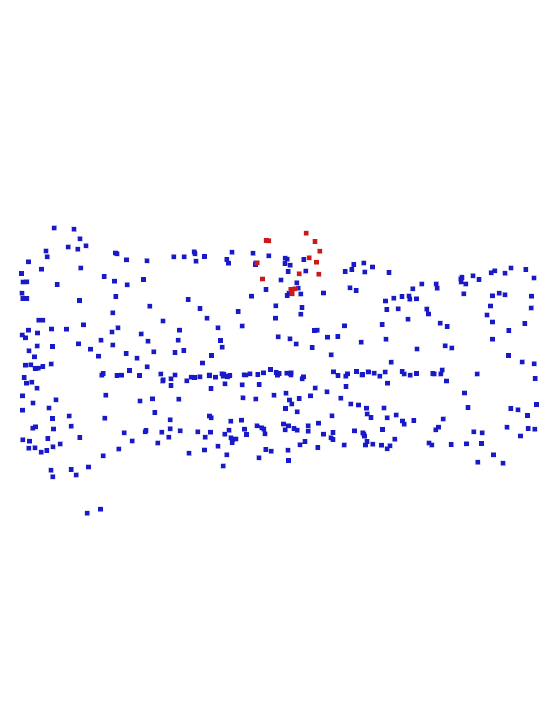}
		\includegraphics[width=.1\linewidth]{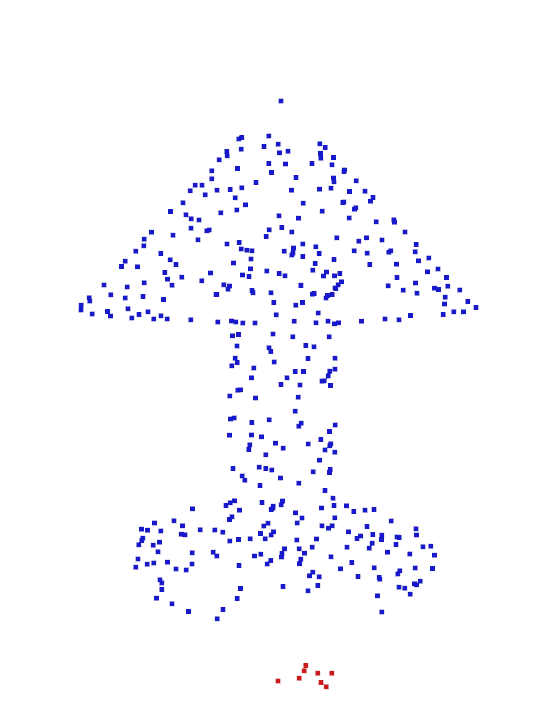}
		\includegraphics[width=.1\linewidth]{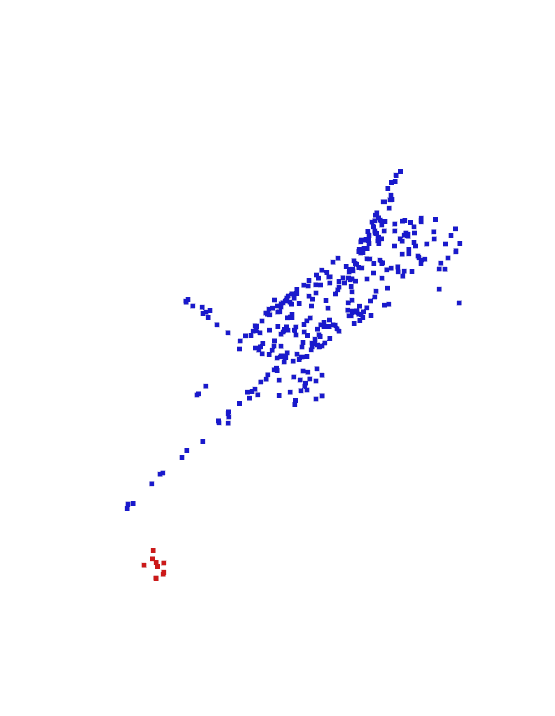}
		\includegraphics[width=.1\linewidth]{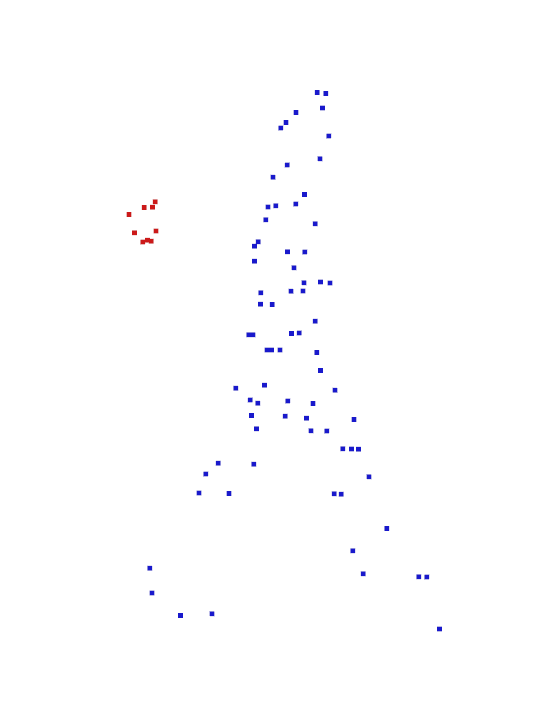}
		\includegraphics[width=.1\linewidth]{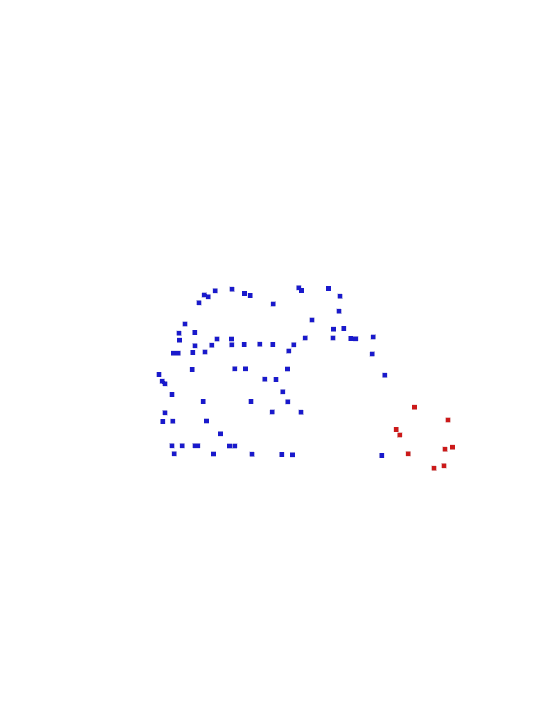}
		\subcaption{Critical points selected by the victim PointNet classifier for the example backdoor training samples in Fig. \ref{subfig:example_bd_sample_RS}.}
	\end{minipage}
	\begin{minipage}[b]{\linewidth}
		\centering
		\includegraphics[width=.1\linewidth]{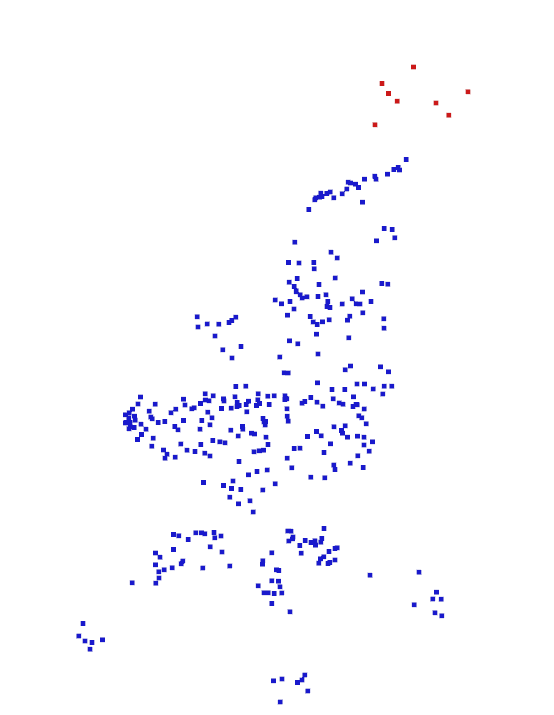}
		\includegraphics[width=.1\linewidth]{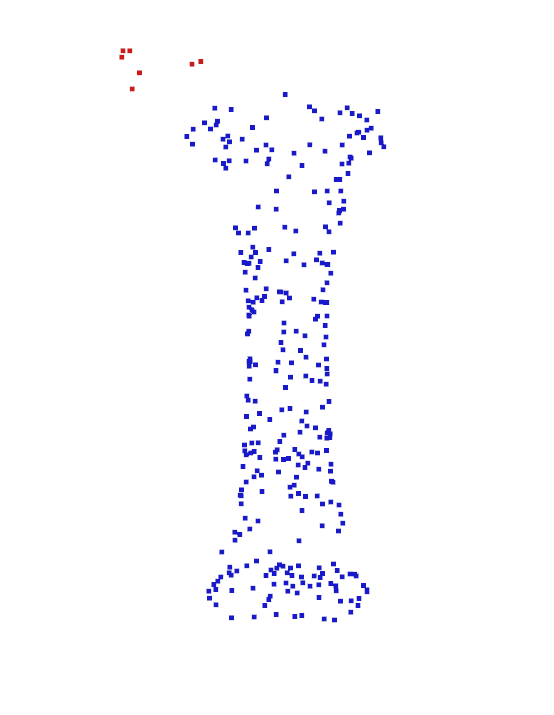}
		\includegraphics[width=.1\linewidth]{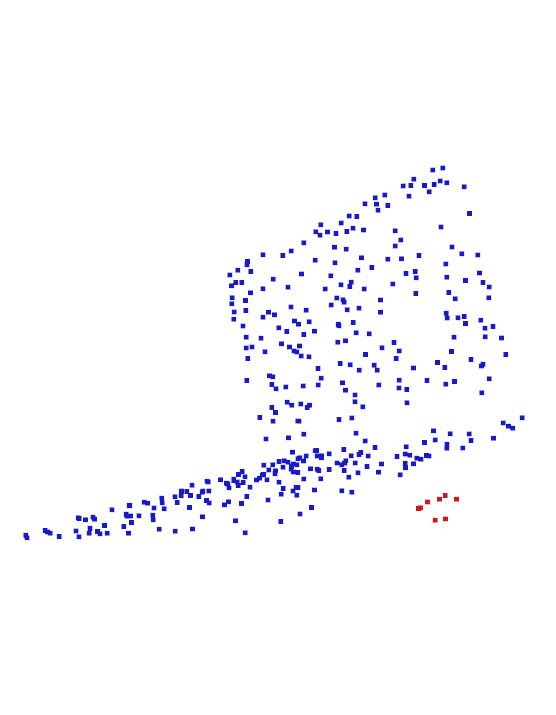}
		\includegraphics[width=.1\linewidth]{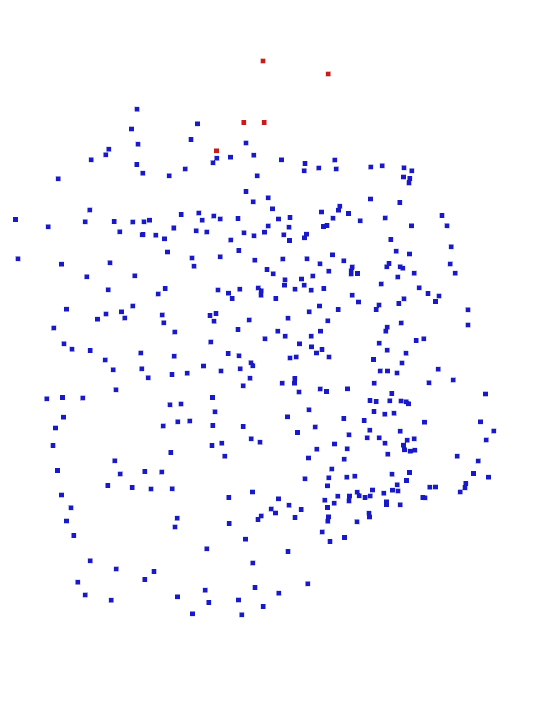}
		\includegraphics[width=.1\linewidth]{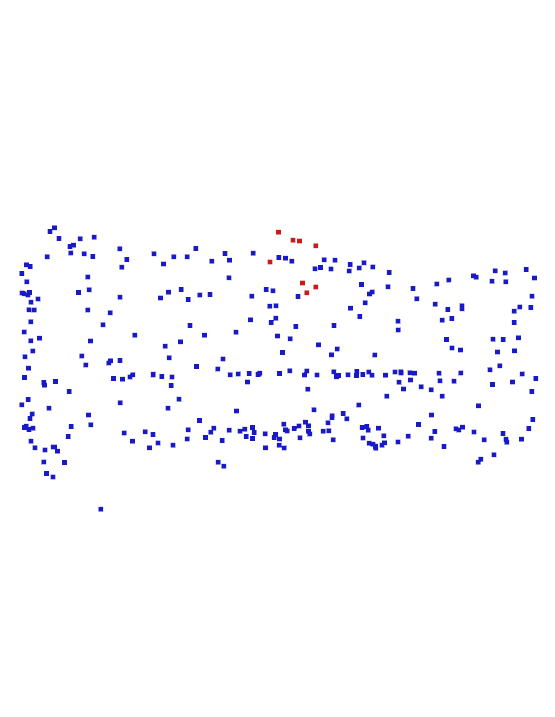}
		\includegraphics[width=.1\linewidth]{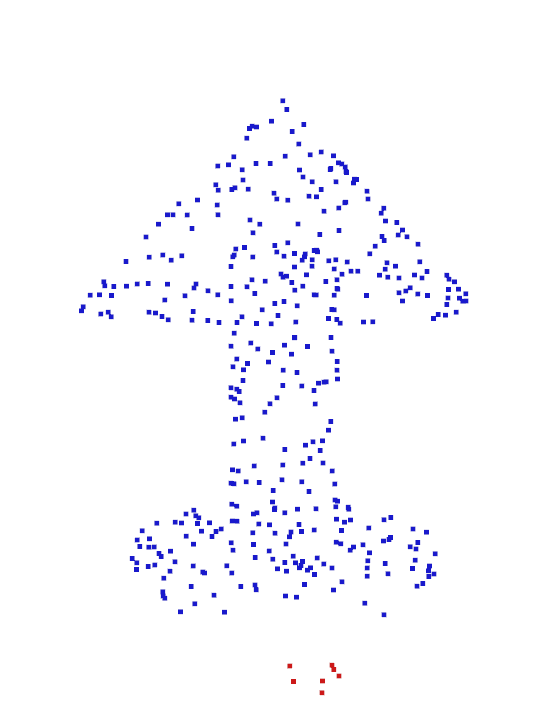}
		\includegraphics[width=.1\linewidth]{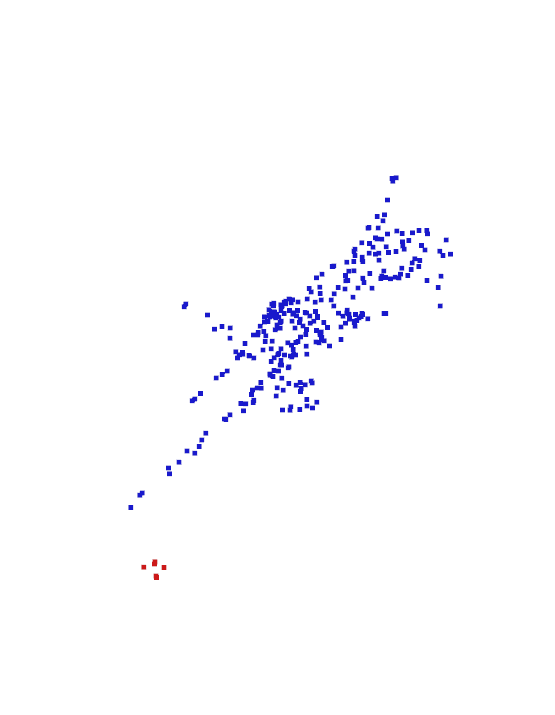}
		\includegraphics[width=.1\linewidth]{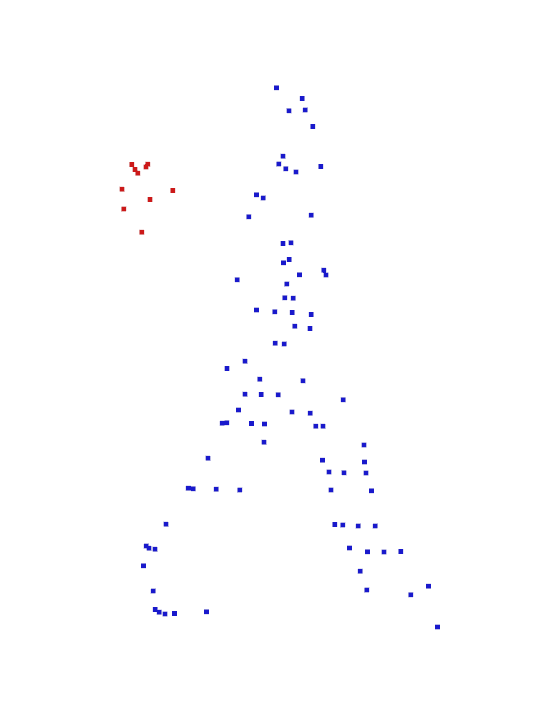}
		\includegraphics[width=.1\linewidth]{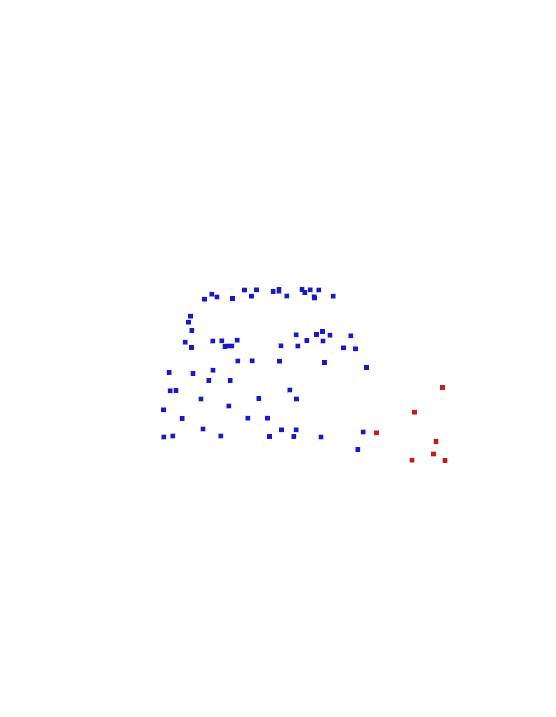}
		\subcaption{Critical points selected by the victim PointNet classifier for the example backdoor training samples in Fig. \ref{subfig:example_bd_sample_RP}.}
	\end{minipage}
	\begin{minipage}[b]{\linewidth}
		\centering
		\includegraphics[width=.1\linewidth]{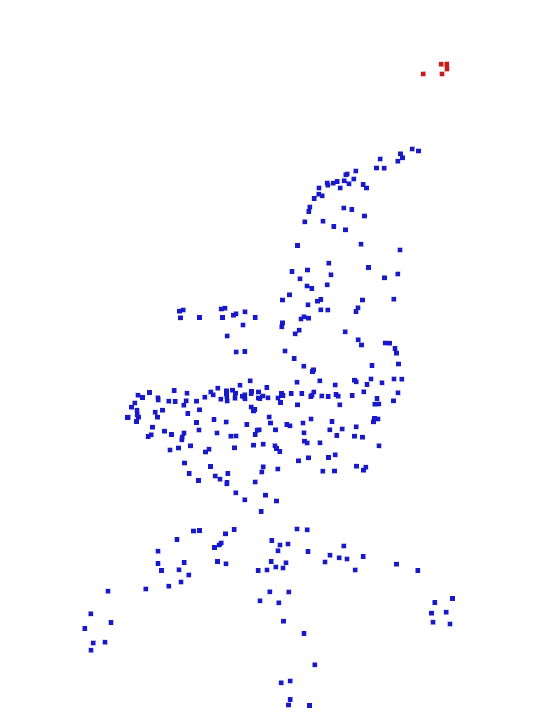}
		\includegraphics[width=.1\linewidth]{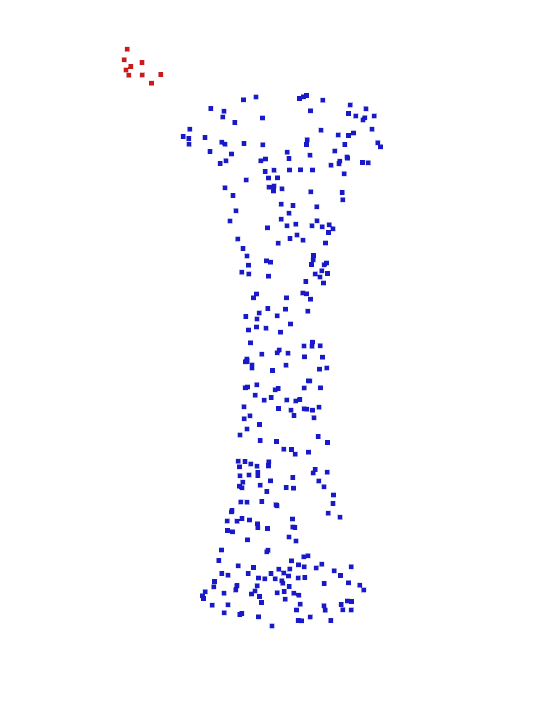}
		\includegraphics[width=.1\linewidth]{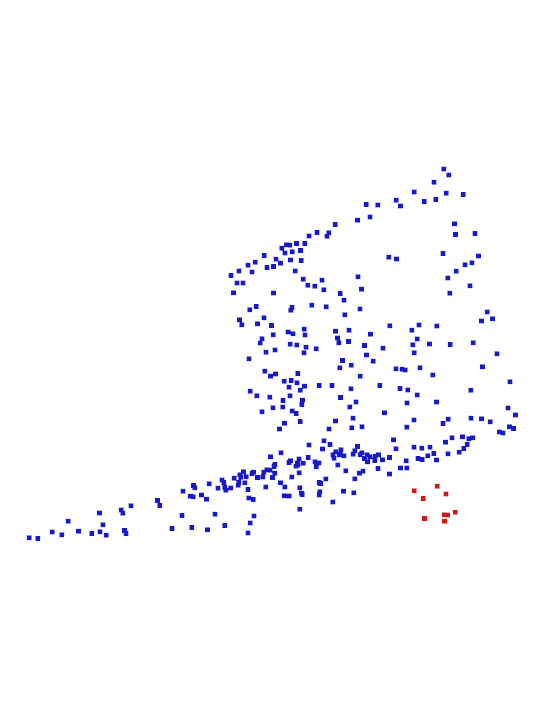}
		\includegraphics[width=.1\linewidth]{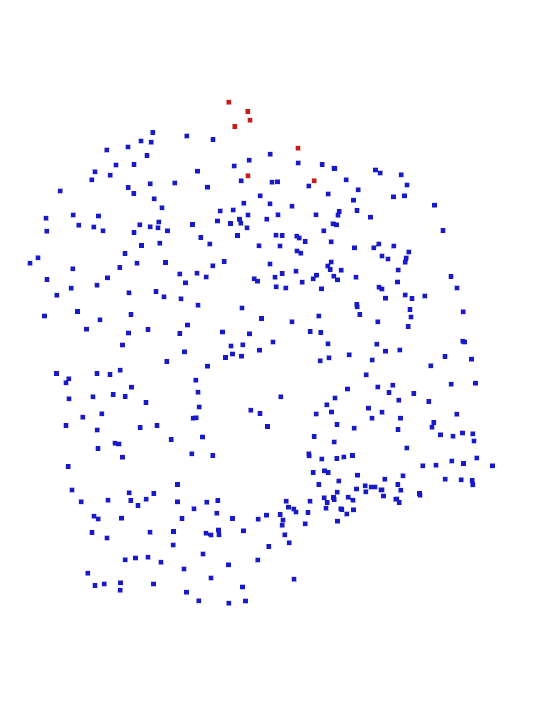}
		\includegraphics[width=.1\linewidth]{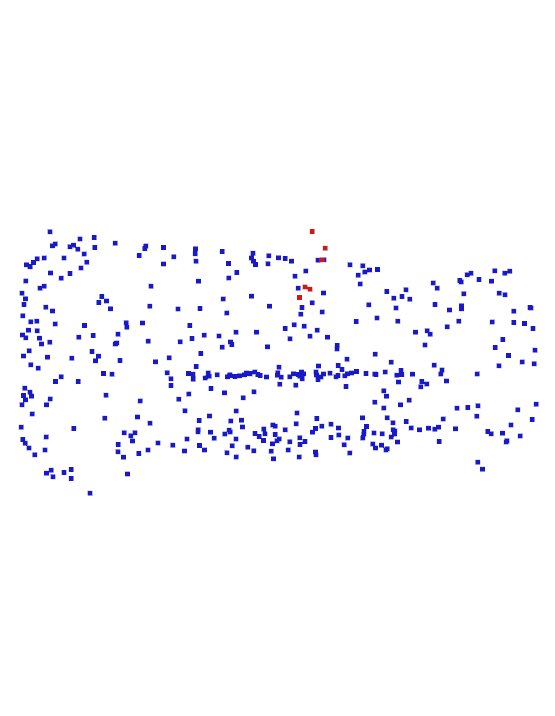}
		\includegraphics[width=.1\linewidth]{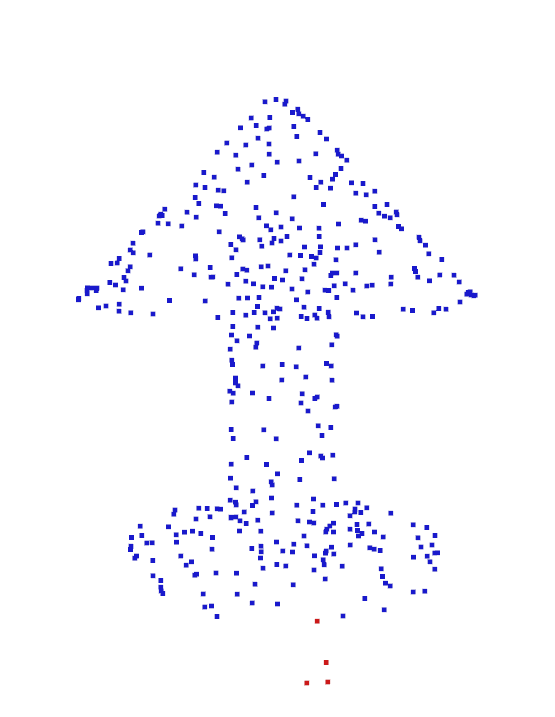}
		\includegraphics[width=.1\linewidth]{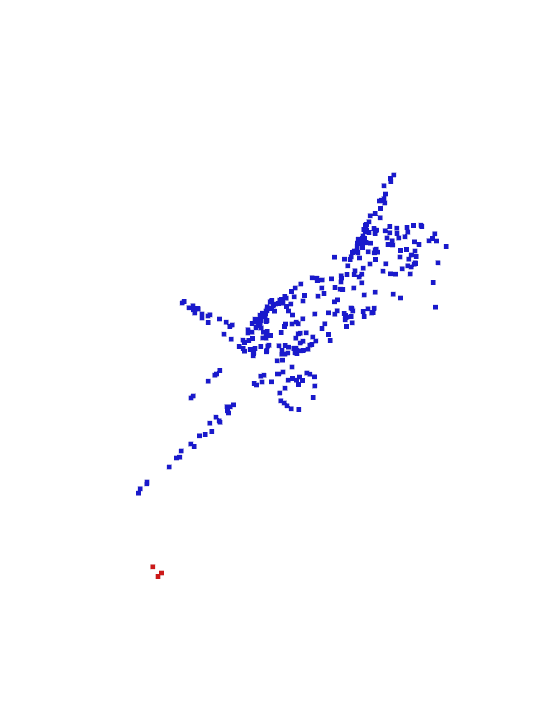}
		\includegraphics[width=.1\linewidth]{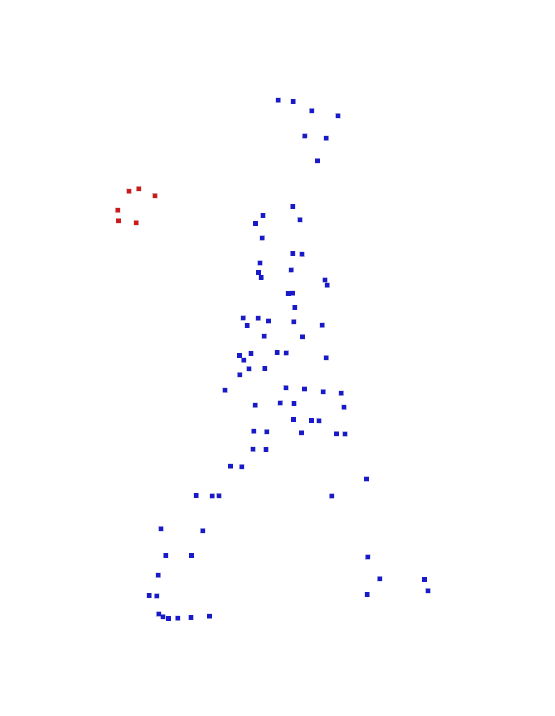}
		\includegraphics[width=.1\linewidth]{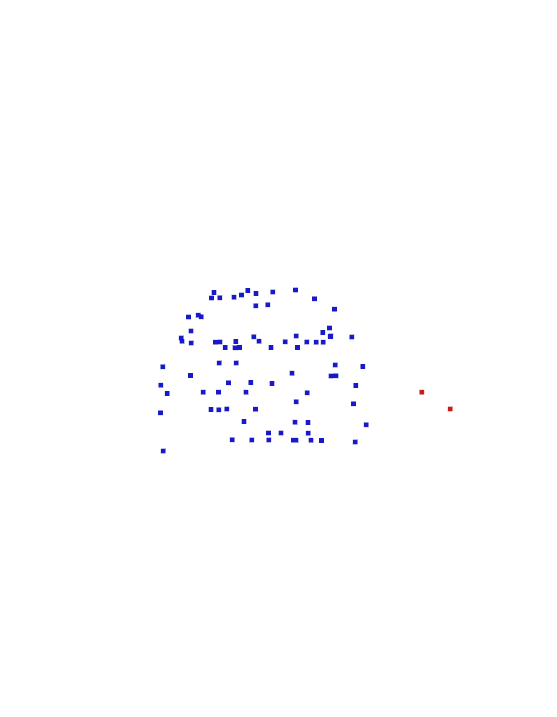}
		\subcaption{Critical points selected by the victim PointNet classifier for the example backdoor training samples in Fig. \ref{subfig:example_bd_sample_HS}.}
	\end{minipage}
	\caption{Critical points selected by the associated victim PointNet classifiers for the example training samples for the 36 attacks. For all PC examples, there is a subset of backdoor points selected as the critical points.}
	\label{fig:example_backdoor_training_sample_critical_points}
\end{figure*}

According to \cite{PointNet}, PointNet selects a subset of points from a PC, using max pooling, as the ``critical points'' that visually represent the skeleton of the object from which the PC is obtained (see Fig. 7 of \cite{PointNet}). To validate that the backdoor mapping has been learned by the victim classifier trained on the poisoned training set, we feed each PC (with backdoor points embedded) in Fig. \ref{fig:example_backdoor_training_sample} into the victim classifier for its associated attack. In Fig. \ref{fig:example_backdoor_training_sample_critical_points}, we show the critical points selected for each PC by the victim classifier with PointNet architecture -- for all 36 PCs, a subset of backdoor points are selected as the critical points; such selection is decisive for misclassification to the attacker's target class.

\end{document}